\documentclass[preprint2]{aastex}

\usepackage{amsmath}
\usepackage{graphicx}
\usepackage{float}
\usepackage[caption = false]{subfig}
\usepackage{enumerate}
\usepackage{amssymb}

\def\SOri{$\sigma$ Ori }
\def\mum{$\mu m$ }
\def\arcmin{$^{\prime}$ }

\slugcomment{A paper to be submitted to ApJ}

\shorttitle{{\em PACS} in Sigma Ori}
\shortauthors{Mauc\'o et al.}

\begin{document}


\title{A Herschel view of protoplanetary disks in the \SOri cluster}


\author{Karina Mauc\'o \altaffilmark{1},
Jes\'us Hern\'andez \altaffilmark{2},
Nuria Calvet \altaffilmark{3}, 
Javier Ballesteros-Paredes \altaffilmark{1},
C\'esar Brice\~{n}o \altaffilmark{4}, 
Melissa McClure \altaffilmark{5}, 
Paola D'Alessio \altaffilmark{1}, 
Kassandra Anderson \altaffilmark{6}, \and
Babar Ali \altaffilmark{7}}

\email{k.mauco@crya.unam.mx}



\altaffiltext{1}{Instituto de Radioastronom\'ia y Astrof\'isica (IRyA), Universidad
Nacional Aut\'onoma de M\'exico (UNAM), Morelia 58089, M\'exico.}
\altaffiltext{2}{Centro de Investigaci\'on de Astronom\'ia (CIDA), M\'erida 5101-A, Venezuela; 
Visiting Scientist, IRyA, UNAM;
Instituto de Astronom\'ia, UNAM, Unidad Acad\'emica en Ensenada, Ensenada 22860, M\'exico.}
\altaffiltext{3}{University of Michigan, UMICH, Ann Arbor, MI 48109}
\altaffiltext{4}{Observatorio Interamericano Cerro Tololo, AURA/CTIO}
\altaffiltext{5}{The European Southern Observatory, ESO}
\altaffiltext{6}{Cornell University, Ithaca, NY}
\altaffiltext{7}{Space Sciences Institute, Boulder, CO}


\begin{abstract}
We present new Herschel PACS observations 
of 32 T Tauri stars in 
the young ($\sim$3 Myr) $\sigma$ Ori cluster.
Most of our objects
are K \& M stars with large excesses at 24 $\mu$m.  
We used irradiated accretion disk models of \citet{dalessio06} to compare
their spectral energy distributions with our observational data. 
We arrive at the following six conclusions.
(i) The observed disks are consistent with irradiated accretion disks systems.
(ii) Most
of our objects (60\%) can be explained by significant dust depletion
from the upper disk layers.
(iii) Similarly, 61\% of our objects can be
modeled with large disk sizes ($\rm R_{\rm d} \geq$ 100 AU).
(iv) The masses of our disks range between 0.03 to 39 $\rm M_{Jup}$, 
where 35\% of our objects 
have disk masses
lower than 1 Jupiter.
Although these are lower limits, high mass ($>$ 0.05 M$_{\odot}$) disks, which are 
present e.g, in Taurus, are
missing. 
(v) By assuming a uniform distribution of
objects around the brightest stars at the center of the cluster,
we found that 80\% of our disks are exposed to external FUV radiation
of $300 \leq G_{0} \leq 1000$, which can be strong
enough to photoevaporate the outer edges of the closer disks.
(vi) Within 0.6 pc from \SOri  
we found forbidden emission lines of [NII] in the spectrum of one of 
our large disk (SO662), but no emission in any of our small ones.
This suggests that this object may be an example of a photoevaporating
disk.\\

\end{abstract}





\section{Introduction \label{sec:int}}

\indent
Due to angular momentum conservation,
the collapse of rotating cloud cores leads
to the formation of stars surrounded by disks.
These disks evolve because they are accreting mass onto
the star and because the dust grains tend to settle towards
the midplane where they collide and grow 
(e.g. Hartmann et al. 1998, Hartmann 2009).
The material
in the disk is subject to irradiation from the
host star and from the high energy fields
produced in accretion shocks on the stellar surface, in
the stellar active regions, and in the environment,
if the star is immersed in the radiation field of
nearby OB stars in a stellar cluster \citep{dalessio01,dalessio06,
adams04,anderson13}. These
high energy fields heat the gas, eventually
leading to its dissipation, while the solids grow to
planetesimal and planet sizes.
Still, many open questions remain on
how these processes happen and interact with each
other.\\
\indent
Previous studies
using Spitzer data of
different star-forming regions
with ages between 1 and 10 Myr,
show a decrease of disk fraction as a function 
of age of the clusters
(Hern\'andez et al. 2007a, henceforward H07a).
The decrease of disk frequency is reflected as a clear drop 
in the mid-IR excess, 
indicating that only 20$\%$ of the stars
retain their original disks by 5 Myr \citep{hernandez07b}.
It is therefore essential to observe disks in the crucial age range between 2 and 10 Myr in which
the agents driving the evolution of protoplanetary disks are most active.
The decrease of the IR excess can be explained by grain growth
and by settling of dust to the disk midplane, reducing the
flaring of the disk and thus its emitting flux. This interpretation is
confirmed by the analysis made by \citet{dalessio06} using 
irradiated accretion disk models. These models simulate the settling process by
introducing the $\epsilon$ parameter, which represents the gas-to-dust mass ratio 
at the disk atmosphere compared to that of the ISM. In this sense, 
a depletion of small grains in the upper layer of the disks 
will be reflected as a small
value of $\epsilon$. Unevolved disks, on the other hand, will have
$\epsilon$ values close to unity.

Once the dust has settled, large bodies
present in the disk will interact with its
local environment creating more complex radial
structures like inner clearings or gaps.
The most probable mechanism responsible for this effect is an orbiting 
companion, either stellar or planetary, that cleared out the material in the inner 
disk \citep{calvet02,espaillat14}.
This mechanism can explain some of the so called transitional and pre-transitional disks (TDs and
PTDs hereafter).  

Also important is disk truncation via mass loss. 
Besides an orbiting planet, truncation of the {\it inner} disk may result 
from the dissipation of gas
being heated by high-energy 
radiation fields
coming from the host star 
\citep{hollenbach00,alexander06b,clarke07,dullemond07}.
Evidence of mass loss in disks
comes from forbidden emission lines of ionized species
like [SII], [OI], [NeII] and [NII].  The low velocity 
component of these lines has been associated with photoevaporative winds
that might be able to explain some of the 
TD and PTDs observed \citep{pascucci09,gorti11}.

The truncation of the {\it outer} parts of the disks, on the other hand,
may be the result of 
environmental effects, like 
mass loss due to
high energy photons from nearby massive
stars impinging on the surface of the disk 
and heating the less tightly bound material.
Expected mass loss rates in externally-illuminated disks 
can be substantial \citep{adams04,facchini16}, 
and when incorporated into viscous evolution models 
\citep{clarke07,anderson13,kalyaan15}
can have a strong impact on the disk structure and lifetime.
Externally-illuminated disks, known as proplyds, have been well 
characterized in the Orion Nebula Cluster
\citep[hereafter ONC;][]{odell94,johnstone98,henney99,storzer99,garcia01,smith05,williams05,eisner08,mann14}
where the radiation from the Trapezium stars 
photoevaporates the disks.
Evidence of outer photoevaporation
in other star-forming regions
has also been found
\citep{rigliaco09,rigliaco13,natta14}.

Multiplicity can also produce truncated disks. 
The fraction of binary companions in young regions 
can be $\sim$30\% or larger where close ($<$ 100 AU) binaries
can affect the evolution of protoplanetary and circumbinary 
disks by significantly reducing their lifetime \citep{daemgen15}.
In Taurus star-forming region the disk population affected
by multiplicity constitutes close binaries with separations $<$40 AU
\citep{kraus12}.

In order to understand what physical processes
cause the disks to evolve, 
many multiband observations of different regions within a
wide range of ages and environments 
have been made. Many of
these studies have used data from the Spitzer Space Telescope
to describe the state of gas and dust within
the first AU from the central object.  
In the $\sim$5 Myr old \citep{dezeeuw99} Upper Scorpius OB association
\citet{dahm09} examined, among others, 7 late-type disk-bearing (K+M) members
using the Infrared Spectrograph (IRS). 
They
found a lack of sub-micron dust grains in the 
inner regions of the disks and
that the strength of silicate emission is spectral-type 
dependent. 
In a disk census performed by \citet{luhman12}
with Spitzer and WISE photometry
they found that late-type members 
have a higher inner disk fraction
than early-types.     
The $\sim$10 Myr old \citep{uchida04} TW Hydrae (TW Hya) association has also been the target for different disk 
evolution studies. 
\citet{uchida04} analyzed two objects with IR excesses 
on their IRS spectra.
They found signs of significant grain growth and dust processing
and also evidence of dust clearing in the inner ($\sim$4 AU) disks,
possibly due to the presence of orbiting planets. 
Similar studies performed in other regions 
like Ophiucus \cite[$\sim$1 Myr]{mcclure10}, Taurus \cite[1-2 Myr]{furlan06}, 
and Chamaeleon I \cite[$\sim$2 Myr]{manoj11}, using IRS spectra
to analyze the strength and shape
of the 10 $\mu m$ and 20 $\mu m$ silicate features, 
have shown that disks in these regions are highly settled 
and exhibit signs of significant dust processing.

In order to describe the distribution of gas and dust in circumstellar
disks around young stars,
many works have been done 
using the Photodetector Array Camera \& Spectrometer (PACS) instrument
on board the Herschel Space Observatory. 
The main idea of these studies
has been the description of disks structures 
as well as the estimation of gas and dust masses
in different star-forming regions
(Riviere-Marichalar
et al. 2013: TW Hya association; Mathews et
al. 2013: Upper Scorpius; Olofsson et al. 2013:
Chamaleon-I).
Additionally, \citet{howard13} 
modeled PACS detections in Taurus,
and found that
the region probed by their observations constitutes the inner part 
(5-50 AU) of their disks.\\
\indent
The $\sim$3 Myr old $\sigma$ Ori cluster (H07a)
is an excellent laboratory
for studies of disk evolution
for two reasons: 
first, the large number of stars still harboring disks allows us to obtain results with statistical significance 
and second, given its intermediate age, one can expect the first traces of disk evolution to become 
apparent.
We present here
new Herschel PACS 70 and 160 $\mu m$ photometry
of 32 TTSs in the cluster, 
with B, V, R and I magnitudes, 2MASS, Spitzer IRAC and MIPS 
photometry from H07a and spectral types
from \citet{hernandez14}, H14 hereafter. 
Our main goal is
to describe the state of the dust on our sample by
analyzing the infrared properties of the stars
and by modeling their SEDs
with irradiated accretion disk models. 
In \S\ref{sec:obs} we describe the observational data and a few 
details about the reduction process; 
in \S\ref{sec:SEDs} we present the SEDs of our objects;
in \S\ref{sec:SOri_pro} we characterize our PACS sources;
our results are shown in \S\ref{sec:results}
where we characterize our PACS disks using spectral indices
(\S\ref{sec:spec_ind}) and modeling the SEDs of individual
objects (\S\ref{sec:modelling}); the discussion is presented 
in \S\ref{sec:disscusion} and
the conclusions are shown in \S\ref{sec:conclusions}. 

\section{PACS Observations \label{sec:obs}}

Our Herschel/PACS imaging survey of the \SOri cluster was 
obtained on March 14, 2012 as part of our Herschel program 
$\rm OT1\_ncalvet\_1$. We used the ``scan map" observational 
template with medium scan speed (20\arcsec/s) to map a square field 
3\arcmin per side. Each scan line was 30\arcmin long, and 134 
overlapping scan lines with a stepsize of 15 \arcsec. The field was 
observed twice at orthogonal scan directions in order to mitigate 
the low-frequency drift of the bolometer timelines. 
We aimed at reaching a 1-$\sigma$ point source sensitivity 
of 2.6 mJy and 6 mJy at 70 {\micron} and 160 {\micron}, respectively.

Our observations were processed using the map-making software
Scanmorphos, which was developed and described by \citet{roussel2013}.
We used  the ``FM6" version of the PACS calibration \citep{balog2014} 
and the data processed with version 9 of the Herschel Interactive Processing Environement 
\citep[HIPE][]{ott2010} software \citep[see ][]{briceno2013}. 
The scales for the 70 {\micron} and 160 {\micron} maps are 
1\arcsec/pixels and 2\arcsec/pixels, respectively. 
Scanamorphos preserves astrophysical emission on all spatial scales, ranging 
from point sources to extended structures with scales just below the map size. 
We performed source detection on the 70 {\micron} and 160 {\micron} maps
using the \texttt{daofind} task in \textsl{IRAF}. 
We extracted aperture photometry for the detected sources 
using \texttt{apphot} in \textsl{IRAF}. Following \citet{fischer13}, 
for the 70 {\micron} images we used an aperture radius of 9.6 arcsec, inner 
sky annulus radius of 9.6 arcsec and sky annulus width of 9.6 arcsec;
for the 160 {\micron} images we used an aperture radius of 12.8 arcsec, 
inner sky annulus radius of 12.8 arcsec and a 12.8 arcsec sky annulus width. 
The photometric error was determined as the sum in quadrature of the measurement error
and the calibration error \citep[see][]{briceno2013}.
The detections were then cross matched with the photometric candidates
selected in H07a. 

\begin{figure}[ht]
\includegraphics[width=1\columnwidth]{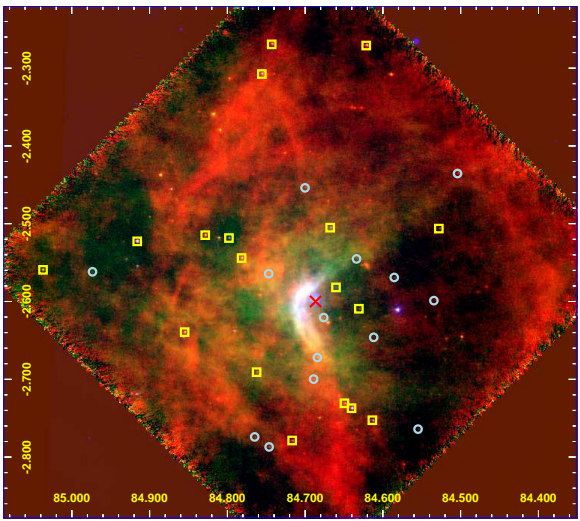}
\caption{Three color-map (red: PACS 160 $\mu$m;
green: PACS 70 $\mu$m; blue: MIPS 24 $\mu$m) 
of the \SOri cluster showing the coverage of PACS observations (big square).
Detections are shown as squares for objects detected at 70 and 160 {\micron} and as circles
for objects only detected at 70 {\micron} (with 160 {\micron} upper limits).
The red cross
indicates the position of the multiple-star system $\sigma$ Ori.
}
\label{fig:PACS_map}
\end{figure}

The PACS field of view (FOV) is shown in Figure~\ref{fig:PACS_map}
as a three-color map using Spitzer MIPS 24 {\micron} (H07a)
and PACS bands.
We detected 32 TTSs in the cluster 
at PACS 70 $\mu m$.
Of these, 17 sources were
also detected at 160 \mum and the rest
only as upper limits.
Stars with 160$\mu$m upper limits were defined using the 3 sigma criteria. 
The large range of values for these upper limits are due to the non-uniform background in the image.
On Figure~\ref{fig:PACS_map} squares indicate stars detected at 70 {\micron} and
160 {\micron} while circles represent stars only detected at 70 {\micron}
(with 160 {\micron} upper limits).
The lowest
flux measured at 70 $\mu$m is equal to 9.4 $\pm$ 1.1 mJy. 
The multiple-star system
\SOri is shown by a red cross on the center of the field, where an arc-shaped nebulosity
can be seen to the west side. This nebulosity consists of gas and dust that have
been dragged away by the strong radiation of the massive stars on the system.

\begin{figure}[ht]
\includegraphics[width=1\columnwidth]{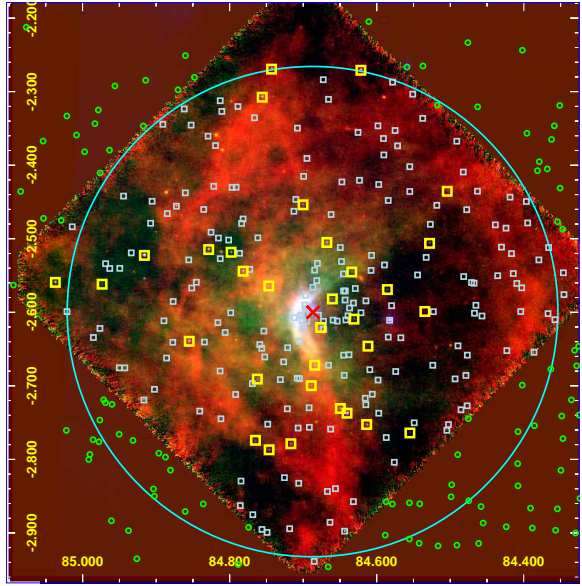}
\caption{Three color-map 
(red: PACS 160 $\mu$m;
green: PACS 70 $\mu$m; blue: MIPS 24 $\mu$m)
of the \SOri cluster showing non detected members 
that lie inside and outside the FOV 
of PACS (lightblue squares and green circles respectively) and 
members with PACS detections (yellow boxes) reported by H14.
The big circle encloses the \SOri 
dense core (extended from the center to a radius of 20$^{\prime}$).
The majority of the members lie inside the FOV of PACS.}
\label{fig:SOri_map}
\end{figure}

The \SOri cluster has a dense core (see Figure~\ref{fig:SOri_map}) extended from the center to a radius
of 20$^{\prime}$, in which most members are located, and a rarefied halo
extended up to 30\arcmin \citep{caballero08}. 
Note that
the dense core is almost entirely covered by the FOV of PACS
that covers a total of 142 TTS of which 23\% are detected at 70 $\mu$m.
These detections have been reported as disk bearing candidates
with infrared excess at 24 {\micron} (H07a). 
The disk fraction
for members inside the dense core is 42\% (H07a).
Our PACS photometry is reported in Table~\ref{tab:observations}.

Others PACS detections not consider in this work were:
(1) 8 sources with optical or 2MASS counterparts detected only at 160 $\mu m$,
all these sources are background candidates from optical-2MASS color magnitude
diagrams; (2) 13 sources detected at 70 $\mu m$ 
(6 of them were reported as non members or galaxy candidates):
SO406 and SO950 are galaxy candidates (Galaxies based on the profile flag of UKIDSS),
SO916 and SO596 are identified as background sources (H07a, H14),
SO457 is a pulsating red giant \citep{caballero10},
SO668 is reported as an obscured QSO at
z = 0.2362 \citep{caballero08},
SO770 is highly

\begin{deluxetable}{cccccc}
\tablewidth{0pt}
\tablecaption{PACS Photometry for Members of the \SOri Cluster \label{tab:observations}}
\tablehead{
\multicolumn{1}{c}{SO ID} & \multicolumn{1}{c}{2MASS ID} & \multicolumn{1}{c}{RA (J2000.0)} & \multicolumn{1}{c}{DEC (J2000.0)} & \multicolumn{1}{c}{70 $\mu$m} 
& \multicolumn{1}{c}{160 $\mu$m}\\ \noalign{\smallskip}
\colhead{} & \colhead{}  & \colhead{(deg)} & \colhead{(deg)} & \colhead{(mJy)} 
& \colhead{(mJy)} \\
}
\startdata
SO299  & 05380097-0226079  &  84.50404  & -2.43553  & 16.8  $\pm$ 1.3   &$<$56.6           \\
SO341  & 05380674-0230227  &  84.52808  & -2.50631  & 42.0  $\pm$ 1.6   & 31.9  $\pm$ 2.7  \\
SO362  & 05380826-0235562  &  84.53442  & -2.59894  & 16.0  $\pm$ 1.3   &$<$16.1           \\
SO396  & 05381315-0245509  &  84.55479  & -2.76414  & 30.5  $\pm$ 1.7   &$<$38.8           \\
SO462  & 05382050-0234089  &  84.58542  & -2.56914  & 26.6  $\pm$ 1.5   &$<$9.8            \\
SO514  & 05382684-0238460  &  84.61183  & -2.64611  & 9.4   $\pm$ 1.1   & ...              \\
SO518  & 05382725-0245096  &  84.61354  & -2.75267  & 71.8  $\pm$ 2.3   & 48.1  $\pm$ 4.9  \\
SO540  & 05382915-0216156  &  84.62146  & -2.271    & 207.0 $\pm$ 5.7   & 243.4 $\pm$ 11   \\
SO562  & 05383141-0236338  &  84.63087  & -2.60939  & 24.0  $\pm$ 1.3   & 18.7  $\pm$ 3.1  \\
SO566  & J053832.13-023243 &  84.63387  & -2.54528  & 16.6  $\pm$ 2.2   &$<$26.7           \\
SO583  & 05383368-0244141  &  84.64033  & -2.73725  & 263.5 $\pm$ 7.2   & 89.1  $\pm$ 5.3  \\
SO615  & 05383587-0243512  &  84.64946  & -2.73089  & 47.0  $\pm$ 1.9   & 93.0  $\pm$ 6.5  \\
SO638  & 05383848-0234550  &  84.66033  & -2.58194  & 23.8  $\pm$ 1.6   &$<$8.2            \\
SO662  & 05384027-0230185  &  84.66779  & -2.50514  & 64.9  $\pm$ 2.7   & 37.4  $\pm$ 2.9  \\
SO682  & 05384227-0237147  &  84.67612  & -2.62075  & 28.6  $\pm$ 2.9   &$<$14.2           \\
SO697  & 05384423-0240197  &  84.68429  & -2.67214  & 21.0  $\pm$ 1.9   &$<$11.7           \\
SO710  & 05384537-0241594  &  84.68904  & -2.69983  & 10.5  $\pm$ 1.4   &$<$17.4           \\
SO736  & 05384803-0227141  &  84.70012  & -2.45392  & 26.8  $\pm$ 1.5   &$<$19.3           \\
SO774  & 05385200-0246436  &  84.71667  & -2.77878  & 50.8  $\pm$ 1.7   & 49.3  $\pm$ 3.7  \\
SO818  & 05385831-0216101  &  84.74296  & -2.26947  & 141.4 $\pm$ 3.9   & 148.9 $\pm$ 6.7  \\
SO823  & 05385911-0247133  &  84.74629  & -2.78703  & 40.9  $\pm$ 1.8   &$<$26.7           \\
SO827  & 05385922-0233514  &  84.74675  & -2.56428  & 31.9  $\pm$ 1.5   &$<$39.8           \\
SO844  & 05390136-0218274  &  84.75567  & -2.30761  & 219.6 $\pm$ 6.0   & 196.6 $\pm$ 10.4 \\ 
SO859  & 05390297-0241272  &  84.76238  & -2.69089  & 22.0  $\pm$ 1.5   & 33.4  $\pm$ 3.3  \\ 
SO865  & 05390357-0246269  &  84.76488  & -2.77414  & 25.8  $\pm$ 1.4   &$<$14.2           \\ 
SO897  & 05390760-0232391  &  84.78167  & -2.54419  & 121.7 $\pm$ 3.6   & 75.8  $\pm$ 7.4  \\ 
SO927  & 05391151-0231065  &  84.79796  & -2.51847  & 100.2 $\pm$ 2.9   & 33.8  $\pm$ 2.5  \\ 
SO984  & 05391883-0230531  &  84.82846  & -2.51475  & 138.8 $\pm$ 3.9   & 179.1 $\pm$ 8.1  \\ 
SO1036 & 05392519-0238220  &  84.85496  & -2.63944  & 107.3 $\pm$ 3.2   & 73.5  $\pm$ 4.5  \\ 
SO1153 & 05393982-0231217  &  84.91592  & -2.52269  & 482.9 $\pm$ 12.8  & 448.6 $\pm$ 19   \\ 
SO1260 & 05395362-0233426  &  84.97342  & -2.56183  & 12.2  $\pm$ 1.1   &$<$31.8           \\   
SO1361 & 05400889-0233336  &  85.03704  & -2.55933  & 199.5 $\pm$ 5.5   & 80.1  $\pm$ 4.5  \\
\enddata
\tablecomments{Column 1:ID following H07a; Column 2: 2MASS ID; Column 3 and 4: right ascension and declination
from H07a; Column 5: PACS 70 $\mu m$ flux; Column 6: PACS 160 $\mu m$ flux.
 }
\end{deluxetable}

\begin{deluxetable}{lccccccccc}
\tablewidth{0pt}
\tablecaption{Properties of 32 PACS sources in the \SOri Cluster \label{tab:properties}}
\tablehead{
\multicolumn{1}{c}{SO ID}  & \multicolumn{1}{c}{Spt} & \multicolumn{1}{c}{$\rm T_{\rm eff}$} & \multicolumn{1}{c}{Av} & \multicolumn{1}{c}{Disk Type} & \multicolumn{1}{c}{$\rm M_{*}$} & \multicolumn{1}{c}{\rm R$_{*}$}  &\multicolumn{1}{c}{$\rm \dot M$} &\multicolumn{1}{c}{Age}
&\multicolumn{1}{c}{$\rm d_{p}$} \\ \noalign{\smallskip}
\colhead{} & \colhead{}  & \colhead{(K)} & \colhead{(mag)} & \colhead{} 
& \colhead{(\rm M$_{\odot}$)} &  \colhead{(\rm R$_{\odot}$)} & \colhead{(\rm M$_{\odot}$ yr$^{-1}$)}
& \colhead{(My)} & \colhead{(pc)}\\
}
\startdata
SO299 & M2.5 $\pm$ 1.0  & 3490.0 &  0.42  & TD & 0.341 & 1.192 & 1.27e-09 & 3.81  &1.54\\
SO341 & M0.0 $\pm$ 0.5  & 3770.0 &  0.58  & II & 0.505 & 1.769 & 1.21e-09 & 2.04  &1.15\\
SO362 & M2.5 $\pm$ 0.5  & 3490.0 &  0.57  & II & 0.348 & 1.641 & 1.04e-09 & 2.16  &0.95\\
SO396 & M1.5 $\pm$ 0.5  & 3630.0 &  0.64  & II & 0.417 & 1.638 & 7.85e-09 & 2.21  &1.32\\
SO462 & M4.0 $\pm$ 1.0  & 3160.0 &  2.94  & II & 0.231 & 2.083 & 1.22e-09 & 1.00  &0.66\\
SO514 & M3.5 $\pm$ 1.0  & 3360.0 &  2.2   & II & 0.254 & 0.885 &   ---    & 6.90  &0.55\\
SO518 & K6.0 $\pm$ 1.0  & 4020.0 &  0.0   & II & 0.754 & 1.367 & 2.88e-09 & 6.50  &1.06\\
SO540 & K6.5 $\pm$ 1.5  & 4020.0 &  0.53  & II & 0.728 & 1.662 &   ---    & 2.99  &2.11\\
SO562 & M3.5 $\pm$ 1.5  & 3360.0 &  0.07  & II & 0.294 & 1.616 & 2.68e-09 & 2.20  &0.35\\
SO566 & M5.0 $\pm$ 1.0  & 2880.0 &  0.59  & II & 0.099 & 1.391 &   ---    & 0.46  &0.47\\ 
SO583 & K4.5 $\pm$ 1.5  & 4330.0 &  0.0   & II & 1.087 & 2.848 & 7.48e-09 & 1.03  &0.90\\
SO615 & K3.0 $\pm$ 3.0  & 4550.0 &  1.98  & EV & 1.498 & 3.019 & 7.7e-10  & 1.46  &0.85\\
SO638 & K2.0 $\pm$ 1.0  & 4760.0 &  1.03  & EV & 1.867 & 3.170 &   ---    & 1.93  &0.20\\
SO662 & K7.0 $\pm$ 1.0  & 3970.0 &  1.89  & II & 0.660 & 2.196 & 1.99e-09 & 1.62  &0.60\\
SO682 & M0.5 $\pm$ 1.0  & 3770.0 &  0.67  & II & 0.505 & 1.785 & 2.3e-10  & 2.00  &0.14\\
SO697 & K6.0 $\pm$ 0.5  & 4020.0 &  0.43  & II & 0.725 & 1.900 & 1.07e-09 & 2.42  &0.45\\
SO710 & M1.5 $\pm$ 1.5  & 3630.0 &  0.59  & II & 0.417 & 1.677 & 1.02e-09 & 2.12  &0.62\\
SO736 & K6.0 $\pm$ 0.5  & 4020.0 &  0.88  & II & 0.748 & 3.528 & 2.38e-09 & 0.69  &0.92\\
SO774 & K7.5 $\pm$ 1.0  & 3970.0 &  0.12  & II & 0.674 & 1.730 & 1.21e-09 & 2.69  &1.13\\
SO818 & M0.0 $\pm$ 0.5  & 3770.0 &  0.54  & TD & 0.509 & 1.346 & 3.4e-10  & 3.84  &2.10\\
SO823 & M2.0 $\pm$ 1.5  & 3490.0 &  2.11  & II & 0.354 & 2.958 &   ---    & 0.38  &1.23\\  
SO827 & M2.5 $\pm$ 1.0  & 3490.0 &  0.0   & II & 0.335 & 1.094 & 7.3e-10  & 4.98  &0.44\\
SO844 & M0.5 $\pm$ 0.5  & 3770.0 &  0.42  & II & 0.506 & 1.754 & 4.38e-09 & 2.08  &1.88\\
SO859 & M2.5 $\pm$ 0.5  & 3490.0 &  0.71  & II & 0.347 & 1.473 & 1.9e-09  & 2.51  &0.74\\
SO865 & M3.5 $\pm$ 1.0  & 3360.0 &  0.0   & II & 0.280 & 1.180 & 7.2e-10  & 3.47  &1.19\\
SO897 & K6.5 $\pm$ 1.5  & 4020.0 &  0.35  & TD & 0.725 & 1.945 & 1.78e-09 & 2.33  &0.69\\
SO927 & M0.0 $\pm$ 1.0  & 3770.0 &  1.04  & II & 0.506 & 1.690 & 9.2e-10  & 2.24  &0.86\\
SO984 & K7.0 $\pm$ 0.5  & 3970.0 &  0.74  & II & 0.665 & 1.983 & 2.83e-09 & 2.08  &1.04\\
SO1036 & K7.5 $\pm$ 0.5  &3970.0 &  0.92  & II & 0.662 & 2.124 & 8.13e-09 & 1.77  &1.08\\
SO1153 & K5.5 $\pm$ 1.0  &4140.0 &  0.15  & I  & 0.875 & 1.402 & 4.19e-09 & 6.99  &1.52\\
SO1260 & M2.5 $\pm$ 1.0  &3490.0 &  0.71  & II & 0.343 & 1.223 & 2.03e-09 & 3.46  &1.81\\
SO1361 & K7.5 $\pm$ 1.0  &3970.0 &  0.53  & II & 0.670 & 1.843 &   ---    & 2.42  &2.21\\
\enddata
\tablecomments{Column 1: ID following H07a; Column 2: spectral type from H14;
Column 3: effective temperature; Column 4: extinction from H14; Column 5: disk type
from H07a; Column 6: stellar mass, Column 7: stellar radius; Column 8: mass accretion rate; 
Column 9: age; Column 10: projected distance to the central \SOri multiple system.
Stellar masses, radii and ages were derived as described in \S\ref{sec:HR_D}  
}
\end{deluxetable}

\begin{deluxetable}{lcccccccc}
\tablewidth{0pt}
\tablecaption{Properties of 74 undetected sources in the \SOri Cluster \label{tab:properties_undetec}}
\tablehead{
\multicolumn{1}{c}{SO ID}  & \multicolumn{1}{c}{Spt} & \multicolumn{1}{c}{$\rm T_{\rm eff}$} & \multicolumn{1}{c}{Av} & \multicolumn{1}{c}{Disk Type} & \multicolumn{1}{c}{$\rm M_{*}$} & \multicolumn{1}{c}{\rm R$_{*}$}  &\multicolumn{1}{c}{$\rm \dot M$} &\multicolumn{1}{c}{Age}\\ \noalign{\smallskip}
\colhead{} & \colhead{}  & \colhead{(K)} & \colhead{(mag)} & \colhead{} 
& \colhead{(\rm M$_{\odot}$)} &  \colhead{(\rm R$_{\odot}$)} & \colhead{(\rm M$_{\odot}$ yr$^{-1}$)}
& \colhead{(My)} \\
}
\startdata
SO117   &  M3.0  $\pm$ 0.5  &  3360  &  0.26  & III   &   0.276   &   1.143   &  7.2E-11    &   3.85\\
SO165   &  M4.5  $\pm$ 1.0  &  3160  &  0.16  & III   &   0.163   &   0.817   &  4.1E-11    &   6.29\\
SO214   &  M2.0  $\pm$ 0.5  &  3490  &  0.57  & III   &   0.35    &   1.855   &  5.5E-10    &   1.77\\
SO219   &  M4.5  $\pm$ 1.5  &  3160  &  1.29  & III   &   0.144   &   0.578   &  8.0E-12    &   9.31\\
SO220   &  M3.0  $\pm$ 0.5  &  3360  &  0.13  & III   &   0.276   &   1.143   &  6.6E-11    &   3.85\\
SO247   &  M5.0  $\pm$ 0.5  &  2880  &  0.0   & II    &   0.075   &   1.063   &  3.45E-10   &   0.83\\
SO271   &  M5.0  $\pm$ 1.0  &  2880  &  0.38  & II    &   0.049   &   0.803   &  1.36E-10   &   4.49\\
SO283   &  M5.5  $\pm$ 0.5  &  2880  &  0.0   & III   &   0.085   &   1.205   &  1.83E-10   &   0.65\\
SO327   &  M4.5  $\pm$ 2.0  &  3160  &  1.77  & II    &   0.144   &   0.578   &  1.33E-10   &   9.31\\
SO397   &  M4.0  $\pm$ 1.0  &  3160  &  0.0   & II    &   0.218   &   1.529   &  1.057E-9   &   2.11\\
SO398   &  M5.5  $\pm$ 1.0  &  2880  &  0.0   & III   &   0.049   &   0.803   &  8.2E-11    &   4.49\\
SO426   &  M3.5  $\pm$ 0.5  &  3360  &  0.6   & III   &   0.254   &   0.885   &  4.2E-11    &   6.90\\
SO432   &  M5.0  $\pm$ 0.5  &  2880  &  0.46  & III   &   0.08    &   1.136   &  1.75E-10   &   0.74\\
SO435   &  M5.0  $\pm$ 0.5  &  2880  &  0.0   & II    &   0.09    &   1.27    &  2.79E-10   &   0.58\\
SO484   &  M4.0  $\pm$ 0.5  &  3160  &  0.42  & III   &   0.178   &   1.001   &  5.0E-11    &   4.53\\
SO485   &  M2.0  $\pm$ 1.0  &  3490  &  1.83  & II    &   0.324   &   0.947   &  8.58E-10   &   6.95\\
SO489   &  M4.5  $\pm$ 0.5  &  3160  &  0.31  & III   &   0.187   &   1.106   &  4.4E-11    &   3.66\\
SO490   &  M4.0  $\pm$ 1.0  &  3160  &  1.14  & II    &   0.191   &   1.156   &  2.073E-9   &   3.28\\
SO500   &  M4.5  $\pm$ 2.0  &  3160  &  0.6   & II    &   0.144   &   0.578   &  1.98E-10   &   9.31\\
SO520   &  M3.5  $\pm$ 0.5  &  3360  &  0.3   & II    &   0.284   &   1.217   &  4.8E-10    &   3.11\\
SO525   &  M3.0  $\pm$ 1.0  &  3360  &  0.39  & III   &   0.295   &   1.669   &  3.23E-10   &   2.10\\
SO539   &  M1.5  $\pm$ 0.5  &  3630  &  0.63  & III   &   0.414   &   1.264   &  1.45E-10   &   3.67\\
SO563   &  K7.5  $\pm$ 0.5  &  3970  &  1.99  & II    &   0.659   &   2.227   &  1.252E-9   &   1.56\\
SO572   &  K6.0  $\pm$ 1.0  &  4020  &  0.91  & III   &   0.726   &   1.867   &  3.03E-10   &   2.50\\
SO582   &  M2.5  $\pm$ 0.5  &  3490  &  0.29  & III   &   0.348   &   1.664   &  1.52E-10   &   2.11\\
SO587   &  M3.0  $\pm$ 1.0  &  3360  &  0.71  & II    &   0.297   &   1.912   &  1.084E-9   &   1.71\\
SO592   &  K7.0  $\pm$ 1.0  &  3970  &  0.25  & III   &   0.665   &   1.983   &  2.66E-10   &   2.08\\
SO598   &  M2.0  $\pm$ 1.0  &  3490  &  2.19  & II    &   0.346   &   1.312   &  9.2E-11    &   2.88\\
SO611   &  K7.0  $\pm$ 1.0  &  3970  &  0.48  & III   &   0.665   &   2.005   &  3.86E-10   &   2.03\\
SO616   &  K7.0  $\pm$ 1.0  &  3970  &  0.2   & III   &   0.665   &   1.983   &  4.86E-10   &   2.08\\
SO624   &  M4.5  $\pm$ 0.5  &  3160  &  0.51  & III   &   0.199   &   1.248   &  3.9E-11    &   2.84\\
SO628   &  M4.5  $\pm$ 1.5  &  3160  &  0.24  & III   &   0.206   &   1.334   &  7.7E-11    &   2.60\\
SO637   &  K6.0  $\pm$ 1.0  &  4020  &  0.78  & III   &   0.718   &   2.191   &  4.28E-10   &   1.84\\
SO646   &  M2.5  $\pm$ 1.0  &  3490  &  1.08  & II    &   0.343   &   1.223   &  7.29E-10   &   3.46\\
SO655   &  M3.5  $\pm$ 0.5  &  3360  &  1.2   & III   &   0.254   &   0.885   &  2.1E-11    &   6.90\\
SO658   &  M4.5  $\pm$ 1.0  &  3160  &  0.18  & III   &   0.168   &   0.883   &  6.0E-12    &   5.63\\
SO669   &  M0.0  $\pm$ 1.0  &  3770  &  0.2   & III   &   0.503   &   1.933   &  4.45E-10   &   1.66\\
SO687   &  M1.0  $\pm$ 1.0  &  3630  &  0.0   & II    &   0.417   &   1.619   &  9.42E-10   &   2.25\\
SO691   &  M1.5  $\pm$ 1.5  &  3630  &  0.25  & III   &   0.418   &   1.823   &  3.5E-10    &   1.81\\
SO696   &  K7.0  $\pm$ 1.0  &  3970  &  0.7   & III   &   0.654   &   2.51    &  1.141E-9   &   1.02\\
SO706   &  G5.0  $\pm$ 2.5  &  5500  &  1.73  & III   &   1.649   &   2.773   &  1.202E-9   &   6.65\\
SO721   &  M5.0  $\pm$ 2.0  &  2880  &  0.0   & III   &   0.085   &   1.205   &  1.73E-10   &   0.65\\
SO723   &  M4.0  $\pm$ 1.5  &  3160  &  0.61  & II    &   0.219   &   1.565   &  1.827E-9   &   2.03\\
SO726   &  M0.5  $\pm$ 1.0  &  3770  &  0.03  & II    &   0.506   &   1.69    &  1.359E-9   &   2.24\\
SO733   &  M1.0  $\pm$ 0.5  &  3630  &  0.35  & II    &   0.417   &   1.474   &  6.48E-10   &   2.60\\
SO740   &  M4.5  $\pm$ 0.5  &  3160  &  0.0   & III   &   0.227   &   1.857   &  1.77E-10   &   1.41\\
SO742   &  M1.5  $\pm$ 1.5  &  3630  &  0.24  & III   &   0.417   &   1.517   &  2.44E-10   &   2.49\\
SO747   &  M0.5  $\pm$ 0.5  &  3770  &  0.59  & III   &   0.501   &   2.109   &  4.54E-10   &   1.29\\
SO748   &  M3.5  $\pm$ 0.5  &  3360  &  0.23  & III   &   0.298   &   2.065   &  4.63E-10   &   1.49\\
SO757   &  M3.0  $\pm$ 1.0  &  3360  &  0.25  & III   &   0.286   &   1.252   &  7.0E-11    &   2.94\\
SO759   &  M3.5  $\pm$ 1.0  &  3360  &  0.19  & EV    &   0.291   &   1.415   &  1.0E-10    &   2.58\\
SO765   &  M3.0  $\pm$ 1.0  &  3360  &  1.66  & III   &   0.287   &   1.286   &  4.2E-11    &   2.86\\
SO785   &  M1.0  $\pm$ 1.0  &  3630  &  0.75  & III   &   0.417   &   1.538   &  2.31E-10   &   2.44\\
SO855   &  M4.0  $\pm$ 1.0  &  3160  &  0.0   & III   &   0.212   &   1.415   &  1.54E-10   &   2.39\\
SO866   &  M4.5  $\pm$ 1.0  &  3160  &  0.15  & II    &   0.163   &   0.817   &  6.3E-11    &   6.29\\
SO879   &  K6.5  $\pm$ 1.0  &  4020  &  0.64  & III   &   0.727   &   1.797   &  2.28E-10   &   2.66\\
SO896   &  M2.5  $\pm$ 1.5  &  3490  &  0.75  & III   &   0.341   &   1.192   &  1.8E-10    &   3.81\\
SO901   &  M4.0  $\pm$ 1.0  &  3160  &  0.05  & EV    &   0.157   &   0.746   &  9.0E-12    &   7.09\\
SO908   &  M3.0  $\pm$ 1.0  &  3360  &  1.51  & II    &   0.289   &   1.32    &  4.0E-10    &   2.79\\
SO914   &  M1.5  $\pm$ 1.0  &  3630  &  0.44  & III   &   0.412   &   1.212   &  6.3E-11    &   4.30\\
SO929   &  K7.0  $\pm$ 1.0  &  3970  &  0.51  & III   &   0.674   &   1.743   &  2.01E-10   &   2.66\\
SO940   &  M4.0  $\pm$ 0.5  &  3160  &  0.55  & III   &   0.183   &   1.055   &  6.0E-12    &   4.08\\
SO947   &  M1.5  $\pm$ 1.0  &  3630  &  0.74  & III   &   0.417   &   1.579   &  4.07E-10   &   2.34\\
SO967   &  M4.0  $\pm$ 0.5  &  3160  &  0.0   & II    &   0.183   &   1.055   &  6.3E-11    &   4.08\\
SO978   &  M1.5  $\pm$ 1.0  &  3630  &  0.44  & III   &   0.405   &   1.073   &  2.2E-11    &   6.15\\
SO999   &  M4.0  $\pm$ 1.0  &  3160  &  1.05  & III   &   0.183   &   1.055   &  1.0E-11    &   4.08\\
SO1000  &  M2.0  $\pm$ 1.0  &  3490  &  0.68  & III   &   0.348   &   1.664   &  2.89E-10   &   2.11\\
SO1017  &  M2.0  $\pm$ 1.0  &  3490  &  0.6   & III   &   0.341   &   1.192   &  1.6E-10    &   3.81\\
SO1027  &  M2.0  $\pm$ 0.5  &  3490  &  0.48  & III   &   0.335   &   1.094   &  1.1E-10    &   4.98\\
SO1052  &  M1.5  $\pm$ 1.0  &  3630  &  0.67  & III   &   0.396   &   0.876   &  3.5E-11    &   9.20\\
SO1133  &  K7.5  $\pm$ 1.0  &  3970  &  0.62  & III   &   0.675   &   1.704   &  2.9E-10    &   2.76\\
SO1207  &  M5.0  $\pm$ 0.5  &  2880  &  0.0   & III   &   0.099   &   1.391   &  2.21E-10   &   0.45\\
SO1250  &  K7.0  $\pm$ 1.5  &  3970  &  0.36  & III   &   0.672   &   1.793   &  2.58E-10   &   2.53\\
SO1268  &  M4.5  $\pm$ 3.0  &  3160  &  1.72  & TD    &   0.151   &   0.667   &  3.6E-11    &   8.07\\
\enddata
\tablecomments{Columns description as in Table~\ref{tab:properties}.}

\end{deluxetable}

\noindent
contaminated by the background 
(located on the arc-shaped nebulosity),
SO848, SO1266, SO1154 and SO1182 have no spectral types 
reported by H14, 
SO1075 has no optical information and
SO1155 is located near the edge of the PACS coverage.
The undetected classical T Tauri stars (CTTSs) have the smallest excesses at MIPS 24 $\mu m$ 
(see section \S\ref{sec:cha_PACS}, Figure~\ref{fig:DCC_MIPS}).

\section{Spectral Energy Distribution \label{sec:SEDs}}

The 32 sources detected by PACS in the \SOri field have stellar counterparts 
that have been characterized by H14. Their spectral types, effective temperatures 
(using the calibrations of Pecaut \& Mamajek 2013), 
and reddening corrections Av are shown in Table~\ref{tab:properties}.

We constructed the spectral energy distributions (SEDs) for all of our targets
using optical B, V, R and I magnitudes (when available), 2MASS, Spitzer IRAC and MIPS photometry 
taken from H07a, and adding 
photometry data from Herschel PACS at 70 and 160 $\mu$m, reported
here. We have added submillmeter data from 
\citet{williams13} for four of our PACS
sources: SO540, SO844, SO984 and SO1153.
The SEDs of our objects are shown in 
Figure~\ref{fig:seds} (filled dots).

Figure~\ref{fig:seds} also shows the \SOri median, 
normalized to the J band of each object,
for stars with PACS detections (solid red line). 
The median values and the first and third quartiles
are in Table~\ref{tab:Median}.
As reference we show the Taurus median estimated from 
photometric data only (dashed line) with quartiles.
To estimate 
the photometric
Taurus median we used J, H and K 2MASS photometry from \citet{hartmann05}, 
Spitzer IRAC and MIPS photometry from \citet{luhman10} and PACS 70, 100 and
160 $\mu m$ photometry from \citet{howard13}, with a total number of 21 stars for the calculation, 
covering a spectral type range from K3 to M4 (Table~\ref{tab:Median_Tau}).
This median was corrected for extinction using the Mathis 
reddening law (Mathis 1990, $R$ = 3.1) with $\rm A_{\rm v}$ from \citet{furlan06}. 
The dotted line represents the median SEDs of non-excess stars
in the \SOri cluster (H07a).

The photometry of the PACS sources was corrected 
for extinction using the Mathis reddening law (Mathis 1990, $R$ = 3.1)
with visual extinctions Av from H14. 
In Figure~\ref{fig:seds} each
panel is labeled with the ID of the source
and its classification, according to the IR excess they exhibit.
We based the classification of stars with different IR excess emission,
in the value of the IRAC SED slope determined from [3.6] - [8.0]
color (n$_{3.6-8.0}$), following H07a. 
In this scheme, class II stars
are defined as systems with n$_{3.6-8.0}$ $>$ $-1.8$ (e.g. SO1036, SO396 and SO562,
where IRAC IR excesses are comparable to the median SED of Taurus); 
stars with IRAC IR excesses above the photosphere and with 
n$_{3.6-8.0}$ $<$ $-1.8$ 
are termed here as evolved (EV) disks (like SO638, where
its IRAC IR excesses fall below both medians); and systems without
any excess emission, below the photospheric limit,
are called class III stars. Objects that are classified 
as class III stars or EV disks by their IRAC IR excess
but that exhibit strong 24 \mum excesses are known as transitional disks, 
which is the case of SO299, SO818 and SO897.

\section{Properties of \SOri sources \label{sec:SOri_pro}}

\subsection{Characterization of PACS Sources \label{sec:PACS_cha}}
\label{sec:cha_PACS}
\indent
As shown in Figure~\ref{fig:SOri_map},
142 TTS in \SOri (H07a) fall in the PACS FOV but only
23\% of them are detected. Here we compare
the detections with the entire TTS population in the field
aiming to understand what makes them different. 

\begin{figure*}[ht]
\centering
\includegraphics[width=0.7\paperwidth]{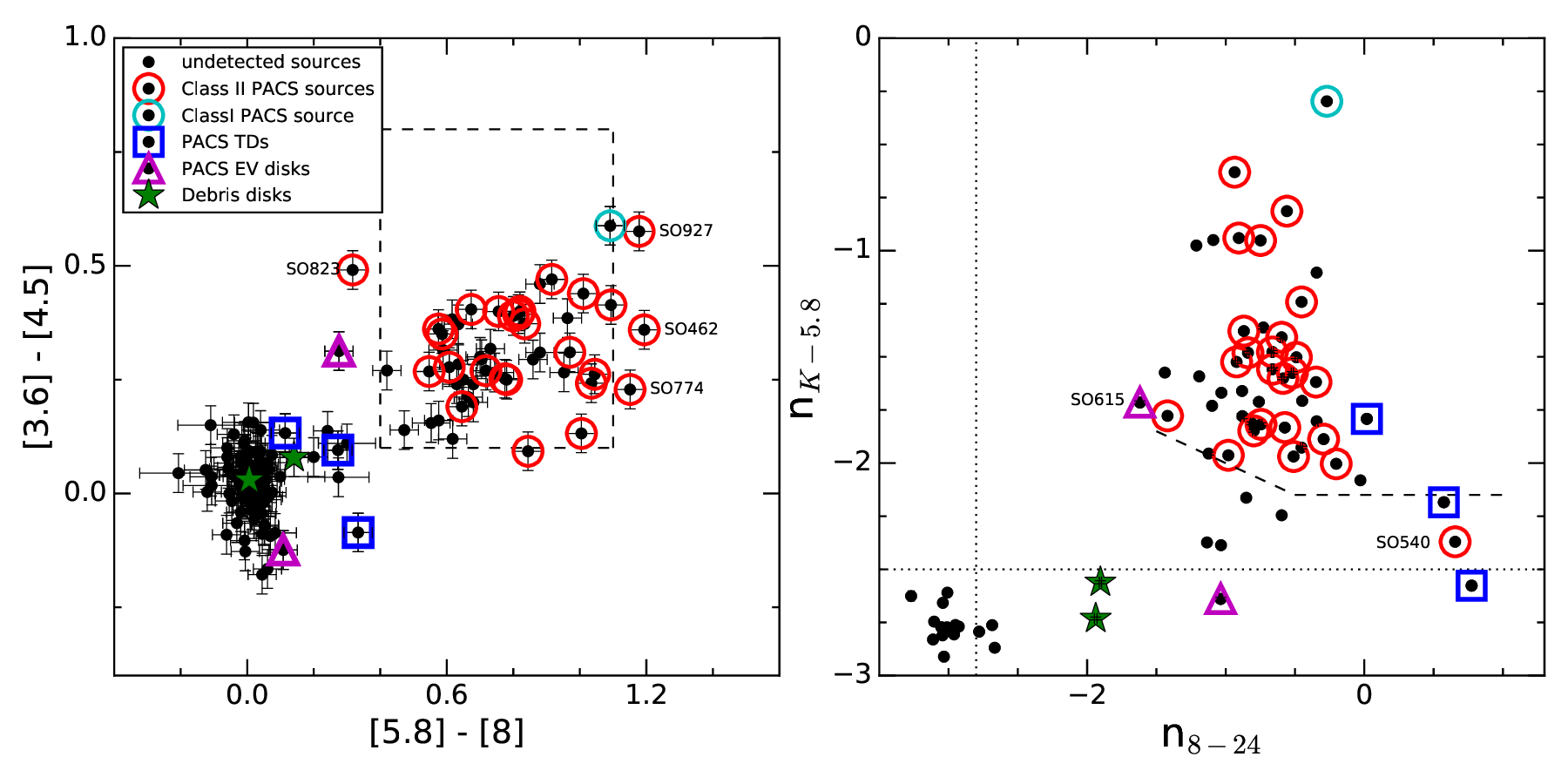}
\caption{{\it Left}: IRAC color-color diagram of members in the \SOri cluster reported by H07a (black dots). 
Different symbols are used to identify PACS detections; circles, squares and triangles 
for classI/class II stars, TD and PTD disks and evolved disks, respectively as classified in H07a.
Green stars indicate the colors of the two known debris disks (DD) in the cluster,
which are not detected by PACS. 
The dashed box encloses the region of predicted
colors for optically thick disks with different accretion rates and different inclinations \citep{dalessio06}.
The object SO823 is described as a slow accretor in H14. The source SO927
has large mid and far IR excesses, but is accreting just above the \citet{barrado03} limit. 
Sources SO462 and SO774 present apparent redder [5.8]-[8.0] colors. 
{\it Right}: SEDs slopes for K-[5.8] and [8.0]-[24] colors
of members in the \SOri cluster reported by H07a. 
Objects detected at PACS are mostly systems
with significant 24 \mum excesses.
The dashed line represents the lower limit of primordial disks in Taurus \citep{luhman10}.
Photospheric limits are indicated by dotted lines.
Based on this diagram SO540
could be a transitional disk candidate instead of a class II star,
and SO615 may be affected by chromospheric contamination
which explains its flat $n_{K-5.8}$ slope (see text for details).
}
\label{fig:DCC_IRAC} 
\end{figure*}

\indent
The left panel of Figure~\ref{fig:DCC_IRAC} shows the PACS 70 $\mu m$ detections in the [3.6]-[4.5] vs [5.8]-[8.0] diagram, as well as
all the \SOri sources in the PACS field.
Sources around [0,0] are referred
as diskless stars while PACS detections are identified following the 
disk classification of H07a.
Note that most of the PACS detected sources are inside the rectangle that encompasses the
optically thick disks region \citep{hartmann05,dalessio06} and are considered CTTSs.
This is consistent with the shape of their SEDs and
their similarities with the Taurus median (Figure~\ref{fig:seds}).
Excluding the TD/PTDs and the two evolved disks in our sample,
which have a decrease of IR excess on the
IRAC bands and therefore will have lower colors than colors predicted by
optically thick disks,
four sources fall outside the loci of CTTSs: SO823, SO462, SO774, and SO927.
The source SO823 
has a low [5.8]-[8.0] color. The rest (SO462, SO774 and SO927) present higher [5.8]-[8.0] colors.
The stars SO462 and SO774 seem to have a slightly decay on the first three IRAC bands 
(3.6, 4.5, 5.8 $\mu$m) which cause an apparent redder [5.8]-[8.0] color 

\begin{figure*}
\centering
\includegraphics[width=0.7\paperwidth]{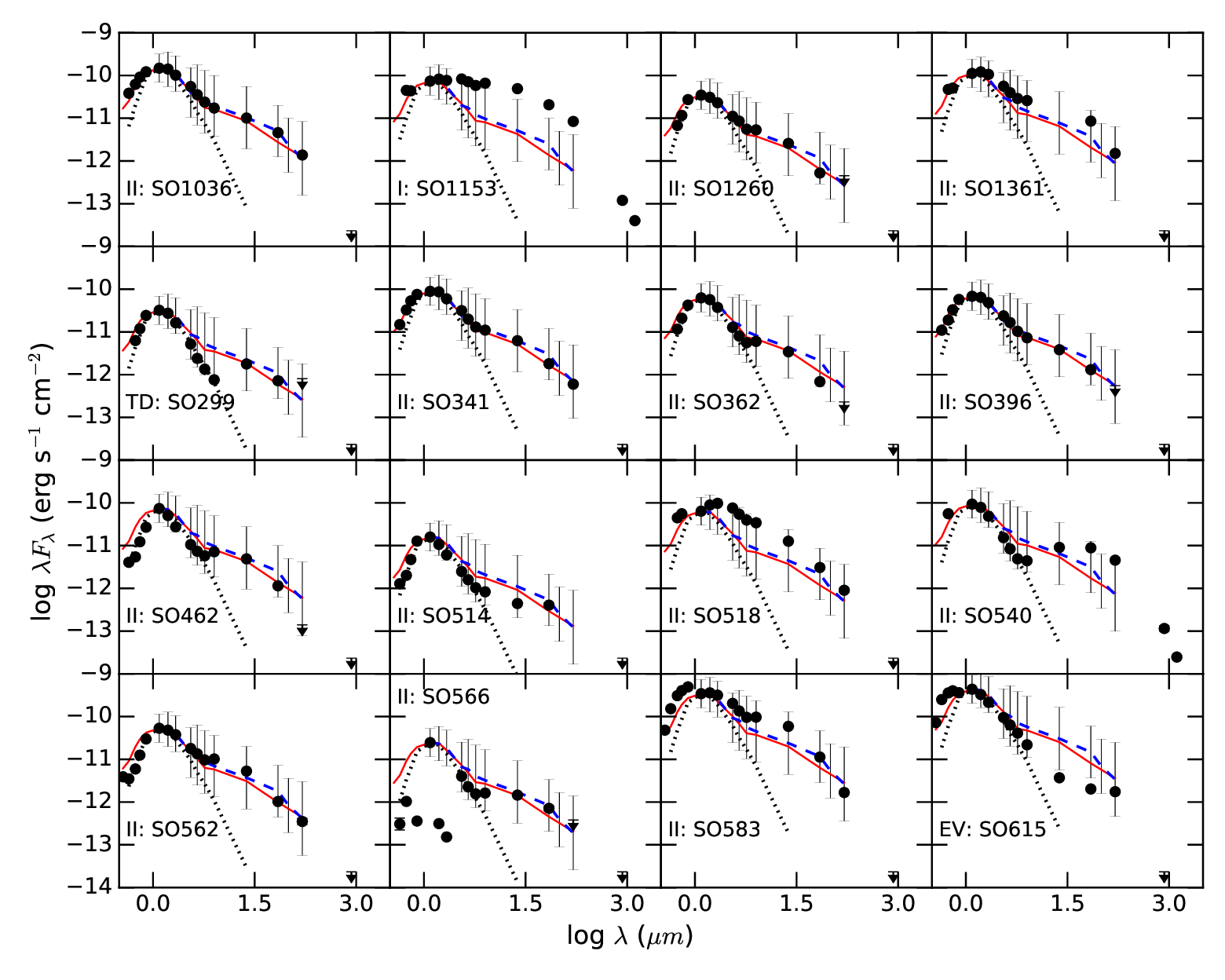}
\caption{Extinction corrected SEDs (filled dots) of PACS sources in the \SOri cluster, 
following the Mathis reddening law according
to the $\rm A_{\rm v}$ reported in H14. 
Each panel is labeled with the ID of the source and
its classification following H07a.
The solid red line represents the \SOri median of PACS sources normalized 
to the J band of each object (Table~\ref{tab:Median}). Also shown is 
the Taurus photometric median 
(blue dashed line) with quartiles, estimated on this work (see text for details). 
The dotted line corresponds to the median photosphere-like fluxes 
in the \SOri cluster (H07a). Error bars are included, 
but in most cases are smaller than the symbol. Downward arrows
indicate upper limits.}
\label{fig:seds}
\end{figure*}

\begin{figure*}
\ContinuedFloat
\centering
\includegraphics[width=0.7\paperwidth]{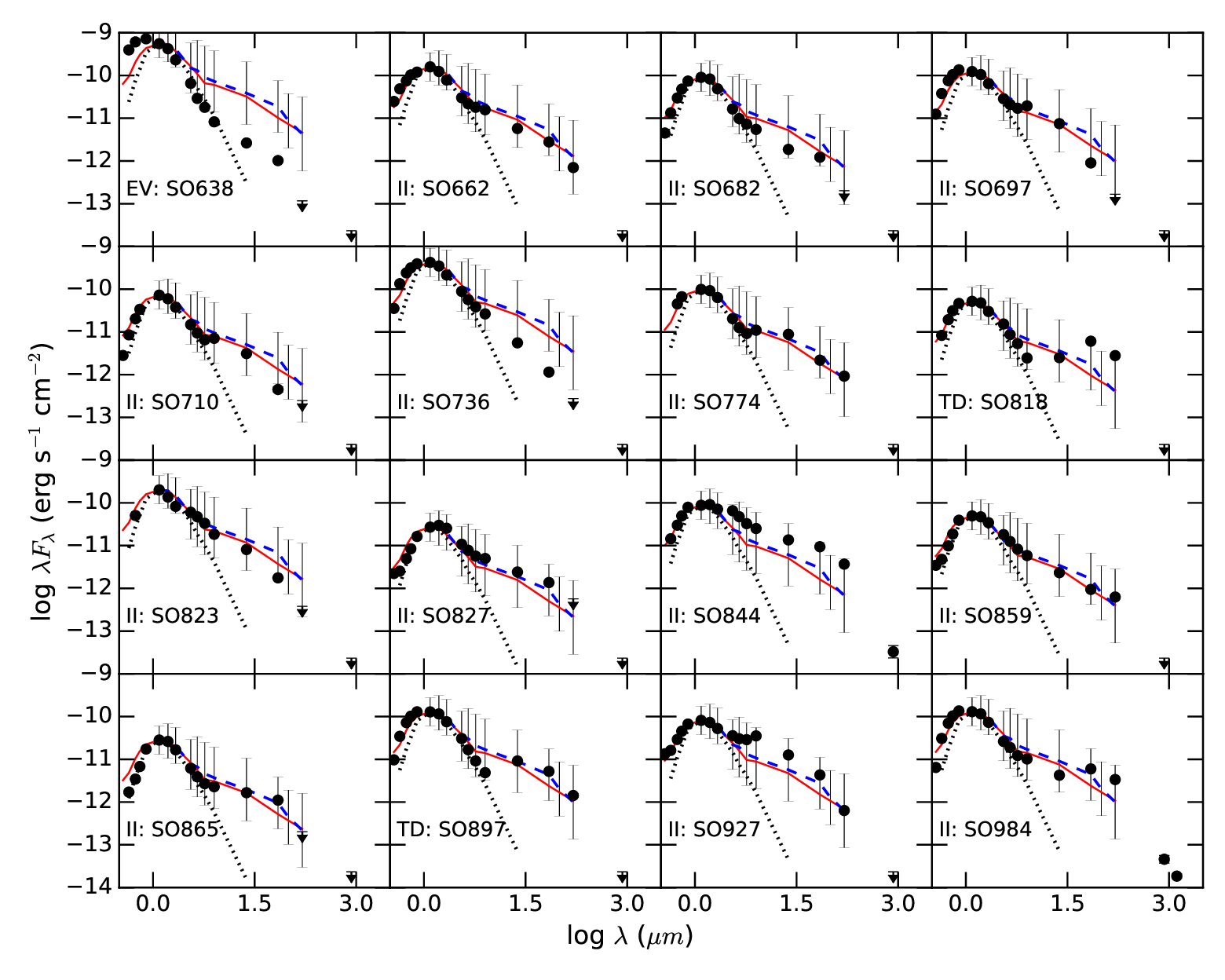}
\caption[]{caption (continued)}
\end{figure*}

\begin{deluxetable}{lccc}
\centering
\tablewidth{0pt}
\tablecolumns{4}
\tablecaption{Median SEDs and Quartiles of Disk Bearing Stars with PACS detections in \SOri \label{tab:Median}}
\tablehead{
\multicolumn{1}{c}{Wavelength} & \multicolumn{3}{c}{log $\lambda$F$_{\lambda}$}\\  \cline{2-4}\noalign{\smallskip}
\colhead{($\mu$m)} & \colhead{Median} & \colhead{Lower} & \colhead{Upper}
}
\startdata
0.44    & -10.84    & -11.35    & -10.42 \\
0.55    & -10.50    & -11.04    & -10.19 \\
0.64    & -10.33    & -10.77    & -10.03 \\
0.79    & -10.18    & -10.56    & -9.89  \\
1.235   & -10.07    & -10.28    & -9.89  \\
1.662   & -10.08    & -10.32    & -9.93  \\
2.159   & -10.25    & -10.48    & -10.10  \\
3.6     & -10.60    & -10.84    & -10.24  \\
4.5     & -10.78    & -11.07    & -10.44  \\
5.8     & -11.00    & -11.25    & -10.54  \\
8.0     & -11.04    & -11.28    & -10.70  \\
24.0    & -11.30    & -11.59    & -11.04  \\
70.0    & -11.81    & -12.00    & -11.32  \\
160.0   & -12.15    & -12.45    & -11.84  \\

\enddata

\end{deluxetable}

\begin{deluxetable}{lccc}
\centering
\tablewidth{0pt}
\tablecaption{Photometric Median SEDs and Quartiles of Disk Bearing Stars in Taurus \label{tab:Median_Tau}}
\tablehead{
\multicolumn{1}{c}{Wavelength} & \multicolumn{3}{c}{log $\lambda$F$_{\lambda}$}\\  \cline{2-4}\noalign{\smallskip}
\colhead{($\mu$m)} & \colhead{Median} & \colhead{Lower} & \colhead{Upper}
}
\startdata

1.235   & -9.07    & -9.25    & -8.92  \\
1.662   & -9.09    & -9.34    & -8.94  \\
2.159   & -9.20    & -9.50    & -9.08  \\
3.6     & -9.64    & -9.94    & -9.36  \\
4.5     & -9.70    & -10.09    & -9.38  \\
5.8     & -9.86    & -10.21    & -9.48  \\
8.0     & -9.95    & -10.25    & -9.54  \\
24.0    & -10.23    & -10.61    & -9.88  \\
70.0    & -10.54    & -10.85    & -10.25  \\
100.0   & -10.87    & -11.04    & -10.40  \\
160.0   & -11.18    & -11.33    & -10.47  \\

\enddata

\end{deluxetable}

\noindent
(note their high 8 $\mu m$ fluxes resembling that of TDs/PTDs in our sample). 
The source SO927 has a peculiar SED with very large near-IR and mid-IR 
excesses. \\
\indent
The right panel in Figure~\ref{fig:DCC_IRAC} displays
the K-[5.8] vs
[8.0]-[24] SED slopes. 
The slope of the SED 
is defined as:
\begin{equation}
 n = \frac{\log{(\lambda_{1} F_{\lambda_{1}})}-\log{(\lambda_{2} F_{\lambda_{2}})}}{\log{(\lambda_{1})}-\log{(\lambda_{2})}}, 
\end{equation}
\noindent
where $\lambda_{i}$ and $\rm F_{\lambda_{i}}$ are the wavelength and the observed flux
at that particular wavelength, respectively. This parameter has been used to classify 
YSOs (Lada et al. 1987, 2006; H07a).
We also indicate
the lower limit of primordial disks in Taurus \citep[][dashed line]{luhman10}.
Photospheric limits are shown by dotted lines.
H07a have classified the object SO540 as a class II star
based on its slopes between 3.6 and 8.0 $\mu m$, however 
using the K-[5.8] and the [8]-[24] slopes indicate
that it could be a TD candidate. 
Another interesting point to notice 
is the fact that one of our two evolved disks, the source SO615, 
appears just in the limit of primordial disks and has the lowest
[8.0]-[24] color of the PACS sample.
This may be the result of chromospheric contamination, 
which is more significant in evolved stars with low mass accretion rates, 
where magnetic activity from the chromosphere of the star 
may produce an extra heating of dust which is reflected
at 5.8 $\mu$m \citep{laura13}.    

\begin{figure*}
\centering
\includegraphics[width=0.7\paperwidth]{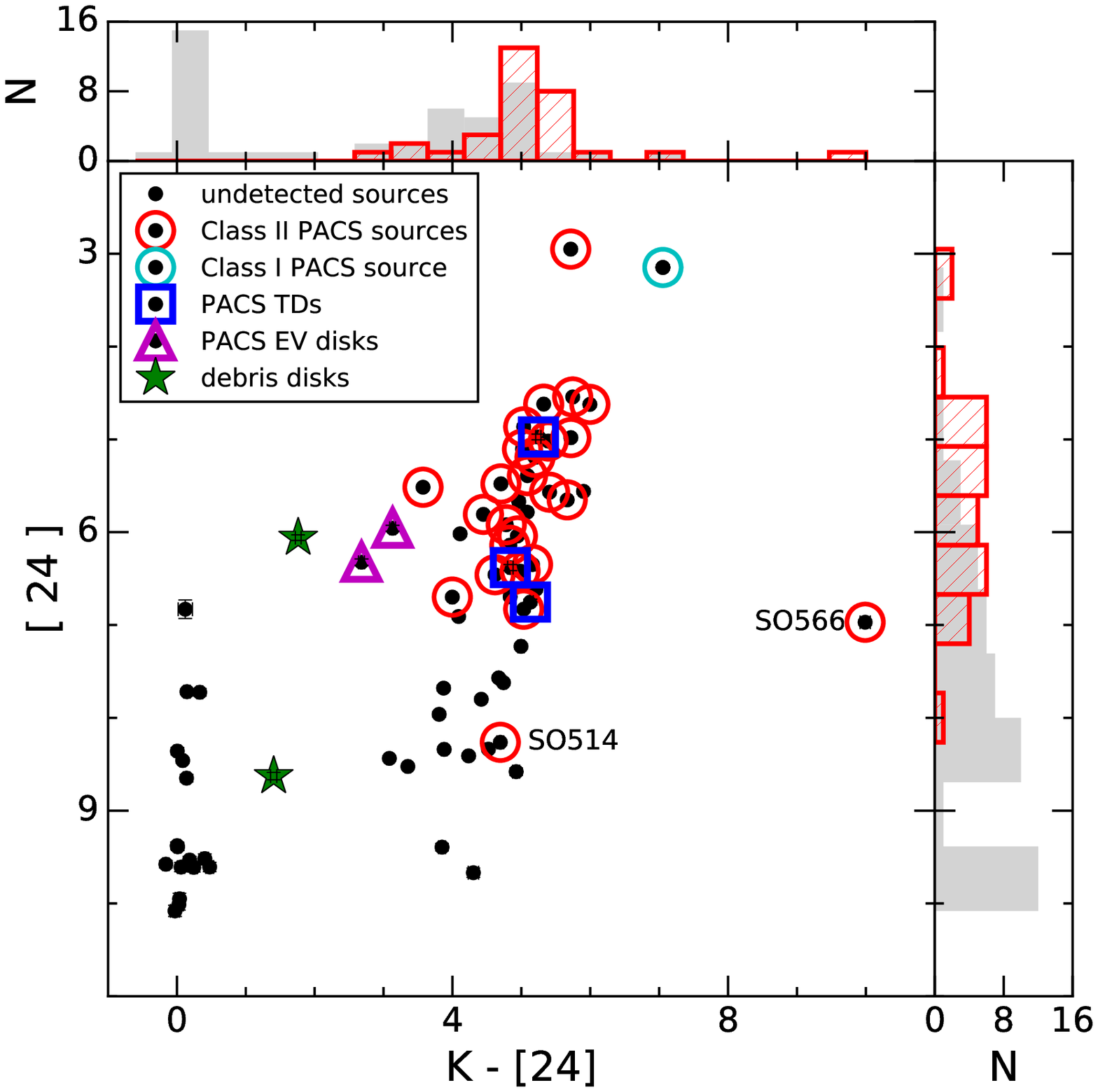}
\caption{[24] vs K - [24] color-magnitude diagram for members in the \SOri cluster. 
Symbols are similar to Figure~\ref{fig:DCC_IRAC}. Also shown are the histograms
of MIPS 24 $\mu m$ (right) and  K - [24] color (top) for detected (red-dashed bars) and non detected (solid gray)
sources. Note that PACS detections are stars with the largest excesses at 24 $\mu m$.
The source
SO566 is a variable star (see Figure~\ref{fig:seds}) thus,
its large K-[24] color may not be real.}
\label{fig:DCC_MIPS} 
\end{figure*}

\indent 
Figure~\ref{fig:DCC_MIPS} shows the [24] vs K-[24]  color-magnitude diagram
of \SOri members highlighting our PACS disks. 
Also shown are the histograms of MIPS 24 $\mu m$ (right) and of K-[24] color
(top). 
As shown in the
figure, our detections have the largest excesses at 24 $\mu m$, 
despite their similar near and mid IR colors with those of non detections (Figure~\ref{fig:DCC_IRAC}),
with
an average magnitude of 5.65 that
corresponds to a flux of 67.27 mJy.
The lowest 24 $\mu m$ magnitude detected corresponds to
the source SO514 with a value of 8.3 mag.
Non detections, on the other hand, present
an average magnitude of 8.16.
The source
SO566 is a variable star 
with variability amplitude in J band larger than 2.7 magnitudes (H14) 
as can be seen in its non-contemporaneous SED
in Figure~\ref{fig:seds},
thus, its large K-[24] color is not real.

\begin{figure}[ht]
\includegraphics[width=1\columnwidth]{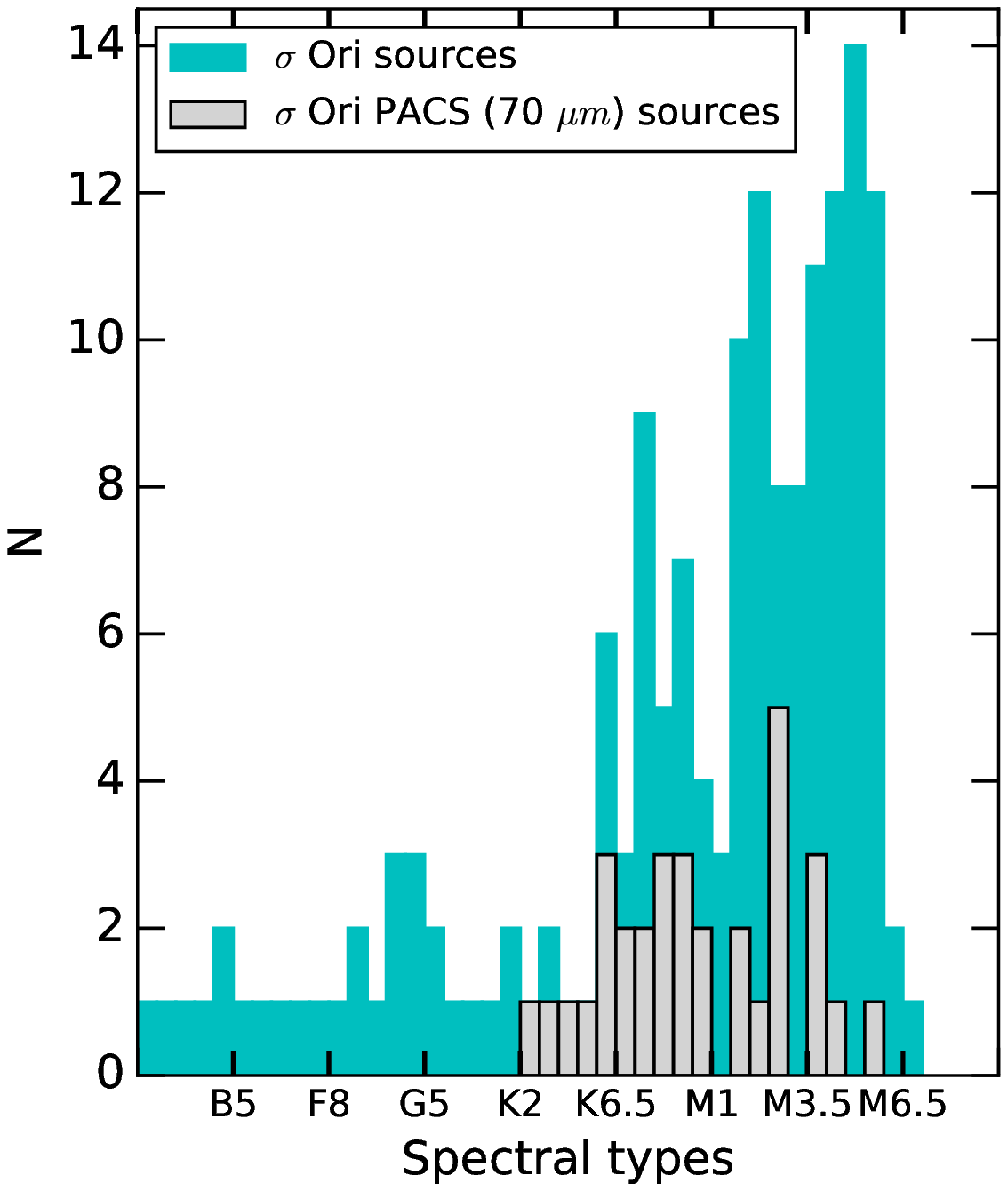}
\caption{Histogram of spectral types reported by H14 for members in the \SOri cluster.
PACS detections are late-type stars with spectral types between K2.0 and M5.0.}
\label{fig:Spt_histo} 
\end{figure}

\indent 
Figure~\ref{fig:Spt_histo} shows an histogram of spectral
types for \SOri members (H14).
PACS 70 $\mu m$ detections cover spectral types from
K2.0 to M5.0. 
Objects with spectral types later than M5 are unlikely to be detected because
the average flux at 70 $\mu$m would be below 
the detection limit for more than a factor of 2 (\S~\ref{sec:obs}).
On the other hand, disk bearing stars with spectral types earlier than K0 
have evolved disks with 24 $\mu$m excesses below the medians of Taurus and $\sigma$ Ori, 
and therefore 70 $\mu$m fluxes below the detection limit,
or are located out of the PACS FOV.

\subsubsection{Transitional and Pre-transitional Disks Statistics \label{sec:TDs}}
\indent
Statistical studies conducted by \citet{muzerolle10} in star forming regions
with ages from $\sim$1 to 10 Myr have shown that
the TDs fraction expected in young regions ($\lesssim$1-2 Myr)
is less that 2\%, while for older regions ($>$3 Myr)
this fraction can be from 7 to 17\%.
However, \citet{kim13} estimated the TDs fraction in the Orion A star-forming region including
PTDs and what they defined as WTDs -intermediate between TDs and PTDs and considered as systems
with an optically thin inner disk separated by a gap from an 
optically thick outer disk- and found that the TDs fraction for clusters with ages $<$1 Myr can be quite high 
(11\% - 25\% for NGC 1333, 21\% - 29\% for ONC and 21\% - 31\% for L1641) 
and similar or greater than the fraction of \citet{muzerolle10} for clusters with ages between 1-2 Myr.
Applying the exact test for the success rate in a binomial experiment 
(R-Statistical Software, Ihaka \& Gentleman 1996), 
we find that the fraction of objects classified
as TDs is 12.5\%, with a 95\% confidence interval  of 0.035 - 0.290. The rather
large uncertainty is due to the modest size of our sample, however, this result is
consistent with the results of \citet{muzerolle10} for a region with an estimated
age of $\sim 3$ Myr.
   
\subsection{Stellar and Accretion Properties \label{sec:HR_D}}

\indent
We estimated stellar and accretion properties of all the 
\SOri members reported in H07a and H14 that lie inside the PACS FOV, for which we
have the necessary spectra and photometry. This information complements 
the spectroscopic census of \SOri sources made by H14.
To characterize the stellar properties of the sources we located them
in the HR diagram.
For this, we
estimated the luminosity of our sample
using J and V 
photometry, visual extinctions $\rm A_{v}$ and spectral types
from H14.
For stars F0 or later,
bolometric correction and effective temperatures
were obtained from the standard table 
for 5-30 Myr old pre-main sequence (PMS) stars from \citet[][table 6]{pecaut13}.
For stars earlier than F0 we used the standard table of
main sequence stars reported by \citet[][table 5]{pecaut13}.
Using these luminosities and effective temperatures we estimated 
stellar radii for \SOri members.
We adopted the Mathis reddening law (Mathis 1990, $R$ = 3.1). 
We assumed a distance to the \SOri cluster of 385 pc \citep{caballero08}.

\begin{figure*}
\centering
\includegraphics[width=0.7\paperwidth]{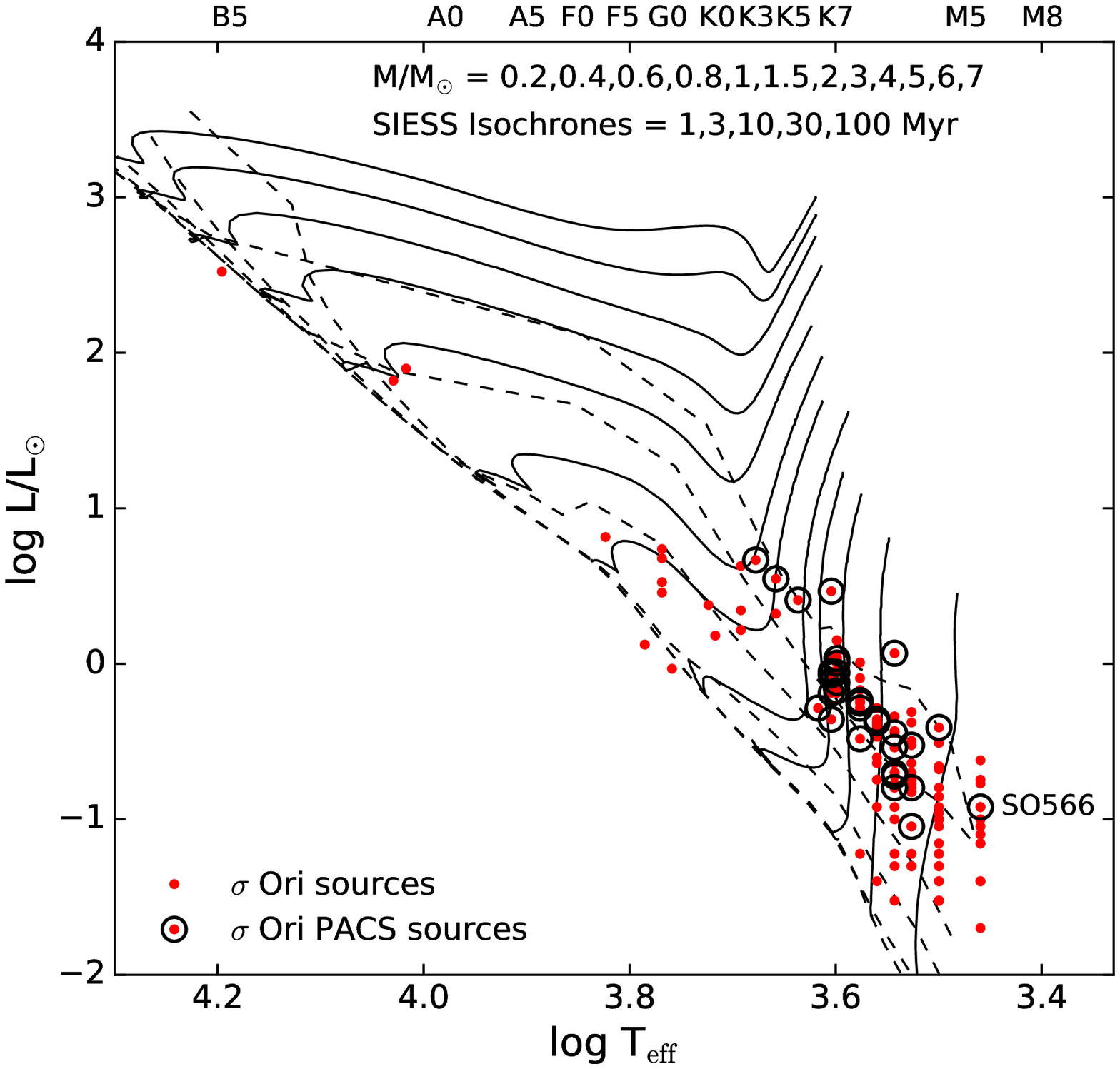}
\caption{HR diagram of \SOri members reported in H07a and H14 with the PACS sources marked. 
The object SO566 is a variable star (see Figure~\ref{fig:seds}) so its position on the HR diagram 
is uncertain.
Solid lines are evolutionary trades for M/$\rm M_{\odot}$ = 0.2, 0.4, 0.6, 0.8, 1, 1.5, 2, 3, 4, 5, 6 and 7
from left to right. Dashed lines are isochrones of 1, 3, 10, 30 and 100 Myr. Evolutionary tracks and
isochrones are from \citet{siess00}.
}
\label{fig:HR_D} 
\end{figure*}

The HR diagram for \SOri members inside the FOV of PACS 
is displayed in Figure~\ref{fig:HR_D}.
Objects detected with PACS are, in general,
late-type PMS stars, as discussed in section~\ref{sec:PACS_cha}. 
Using the SIESS stellar models isochrones \citep{siess00} we were able to calculate the age and mass 
of all our stars in the sample. 
Stellar properties are listed in Tables ~\ref{tab:properties} and ~\ref{tab:properties_undetec}
for detected and undetected sources, respectively.

\indent
Mass accretion rates ($\rm \dot M$)
were estimated from the $\rm H\alpha$ line luminosity 
\citep{muzerolle03,muzerolle05,laura13,rigliaco12,natta04}.
The $\rm H\alpha$ luminosity was estimated
by approximating the flux of the line (F$_{\rm H\alpha}$) as EW$\rm (H\alpha$) x F$_{\rm cont}$,
where F$_{\rm cont}$ and EW$\rm (H\alpha$) are the continuum flux around the line and 
the equivalent width, respectively. We calculated F$_{\rm cont}$ from 
the R$_{\rm c}$ magnitude of each source reported in H07a corrected by 
extinction, and the 
equivalent widths EW(H$\alpha$)
from low resolution spectra reported in H14. Once the lines fluxes are estimated, 
the line luminosities are given by
\begin{equation}
L_{H\alpha} =  4\pi d^{2} F_{H\alpha},
\end{equation}
\noindent
where $d$ is the distance to the \SOri cluster.

\indent
In order to obtain the accretion luminosities we
used the relation between the H$\alpha$ luminosity (L$_{\rm H\alpha}$) and the
accretion luminosity (L$_{\rm acc}$) from \citet{laura13}:
\begin{equation}
\log(L_{acc}) = \rm 1.0(\pm 0.2)\log(L_{H\alpha}) + 1.3(\pm 0.7)
\end{equation}

\indent
Finally, the L$_{\rm acc}$ can be converted into $\rm \dot M$ as
\begin{equation}
 \dot M = \frac{L_{acc} R_{*}}{GM_{*}},
\end{equation}
\noindent
where $\rm M_{*}$ and $\rm R_{*}$ are the stellar mass and radius respectively. 
Three sources have no estimation of mass accretions rates since they do not 
have photometry at the $\rm R_{c}$ band (SO540, SO566 and SO638).

\begin{figure}[ht]
\includegraphics[width=1\columnwidth]{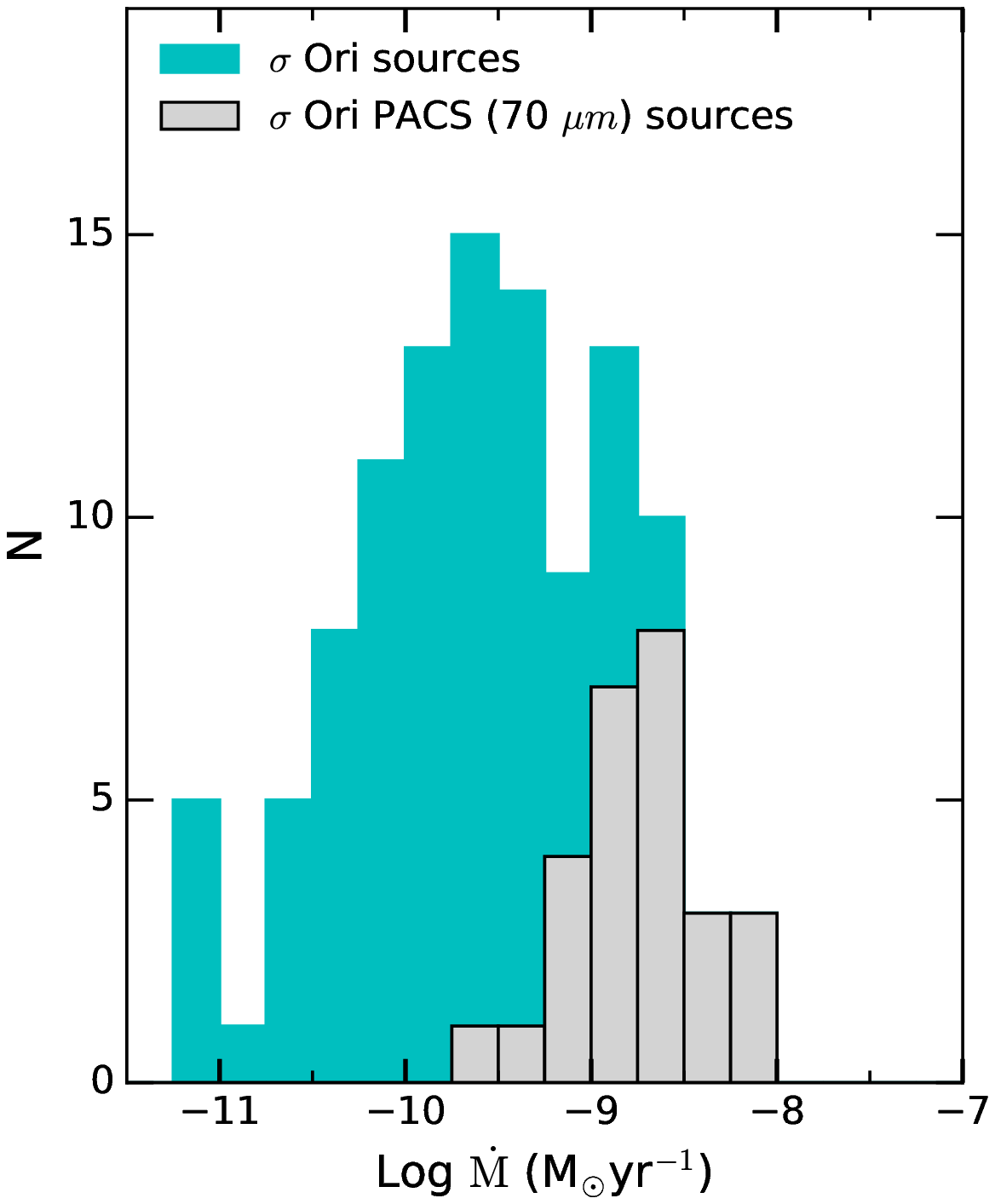}
\caption{Accretion rates of \SOri members. PACS (70 $\mu$m) detections are shown 
as gray bars while undetected members (inside the FOV of PACS) are plotted as light blue bars. 
PACS sources are consistent with accretion rates $\rm \dot M$ $\leq 10^{-8}$ $\rm M_{\odot}yr^{-1}$.
Values of $\rm \dot M < 10^{-9}$ $\rm M_{\odot} yr^{-1}$ 
are unreliable because of chromospheric contamination (see text for details).}
\label{fig:Mdot_histo} 
\end{figure}

Figure~\ref{fig:Mdot_histo} shows the histogram of the mass accretion rates for the \SOri
cluster.
PACS (70 $\mu m$) detections are plotted as gray bars while 
undetected sources as light blue bars.
PACS detections exhibit $\rm \dot M \leq 10^{-8}$ $\rm M_{\odot} yr^{-1}$.
Note, however, that values of $\rm \dot M < 10^{-9}$ $\rm M_{\odot} yr^{-1}$ 
are unreliable because of chromospheric contamination
as demonstrated by \citet{laura13}, where an active chromosphere 
can mask all evidence of an accretion
shock excess for the lowest accretors. 
Additionally, chromospheric activity adds uncertainty to the estimation 
of the luminosities of the lines associated with accretion, 
for low mass stars at later evolutionary stages \citep{manara13}. 
As noted in the figure, 23 undetected sources have $\rm \dot M \geq 10^{-9}$ $\rm M_{\odot} yr^{-1}$,
this is consistent with the discussion on Figure~\ref{fig:DCC_IRAC} and \ref{fig:DCC_MIPS} 
where the undetected CTTSs in the PACS FOV are those with the lowest 
excesses at 24 $\mu$m. Most of these sources constitutes the less massive and 
intrinsically weaker CTTSs in the cluster.

We compared our mass accretion rates with those reported by \citet{rigliaco11},
who estimated $\rm \dot M$ from U band excesses. 
In general, our mass accretion rates are higher than Rigliaco's. 
For the PACS sample we found that 10 out of 19 sources are consistent
with the values reported by \citet{rigliaco11} within a factor
less than 3, while for the undetected sources we got that 
10 out of 16 sources agree within the same factor.
\citet{rigliaco11} neglected extinction, but H07a found extinctions up to
2.94 magnitudes which can be significant at the U band. 
This can explain some of the discrepancies between their values and ours. 
Another source of uncertainty
is due to the fact that mass accretion rates determined by the U band excess
are not adequate for sources with $\rm \dot M$ $<$ $10^{-9}$ $\rm M_{\odot} yr^{-1}$  
because of chromospheric
contamination \citep{laura13}, which is the case of several of our undetected sources. 
Therefore, we also compared our mass accretions rates with
the ones reported by \citet{rigliaco12}, who
estimated $\rm \dot M$ from the H$\alpha$ luminosities.
Even though they also neglected extinction, accretions rates
estimated by $\rm H\alpha$ line luminosities 
are less affected by extinction effects than 
$\rm \dot M$ determined from U band excesses.
We found that,
for the six sources we have in common, all but one (SO490) agree within a factor
$\lesssim$ 3. \citet{gatti08} used near-IR hydrogen recombination lines
(Pa$\gamma$) to measure mass accretion rates of 35 objects
in \SOri. If we compare our mass accretion rates with Gatti's we found that,
for the 25 objects we have in common, all but three agree within a factor $\leq$ 3.
The main uncertainties are due to the stellar mass adopted in each case, 
which in turn depends on the luminosity and the extinction of the sources, 
and on the spectral type.

All the accretion parameters of PACS sources are shown in Table~\ref{tab:properties}.
Additionally, Table~\ref{tab:properties_undetec} lists the accretion
parameters for \SOri members in the PACS FOV but without detections.

\section{Results \label{sec:results}}

\indent
In this section we 
examine the emission of the disks around
the PACS sources in \SOri and interpret it in terms
of irradiated disk models including dust settling. We
first make a global comparison of the spectral
slopes in the PACS range with those predicted by
irradiated disk models.
We then model individual objects to provide
a more detailed characterization of their disks.
This detailed analysis of a fairly large number of disks 
in a young cluster, with estimated stellar and accretion parameters, 
will provide a census of disk 
properties useful for studies of the 
evolution of protoplanetary disks. 

\subsection{Disk Models}
\indent
We follow the methods of 
\citet{dalessio06} to 
calculate the structure and emission of 
accretion disk models.
The main input parameters are 
the stellar properties (M$_{*}$, R$_{*}$, L$_{*}$), 
the mass accretion rate ($\rm \dot M$),
the viscosity parameter ($\alpha$), 
the cosine of the inclination angle ($\mu$),
disk outer radius (R$_{\rm d}$), maximum grain size
at the disk midplane ($\rm a_{maxb}$), and the dust settling; 
assumed constant throughout the disk.

\indent
The composition of the disk consists of amorphous silicates (pyroxenes) 
with a mass fraction of 0.004
and carbonates in the form of graphite with a mass fraction of 0.0025 
(all these mass fractions are relative 
to gas).
The PACS fluxes did not show a strong 
dependence on H$_{2}$O abundance, so we 
kept this parameter fixed at $1e-5$. 
The dust disk inner edge or wall emission is calculated
with the stellar properties, the maximum grain size ($\rm a_{maxw}$),
and the temperature (T$_{\rm wall}$) assumed to be the sublimation
temperature of silicate grains (1400 K) for the diffuse ISM \citep{draine84,dalessio06},
with a dust composition entirely formed by pyroxenes.

\indent
To simulate the settling of dust, \citet{dalessio06} considered the ideal case 
of two dust populations, which differ in grain size. 
Both populations follow
a size-distribution function of the form 
$n( a) \propto a^{-p}$; where $a$ is the radius of the grain, 
whose minimum value is $a_{\rm min}$= 0.005 $\mu m$, and an exponent 
$p$ equal to 3.5, resembling the size-distribution
found in the ISM. The first population corresponds 
to small grains with $a_{\rm max}= 0.25$ $\mu m$-characteristic of ISM dust-
which is expected in unevolved disks. 
These grains are mostly located in the upper layers and are assumed 
to be well mixed with the gas throughout the disk.  
The second population 
consists of larger grains, $a_{\rm max}$= 1 mm, and lives in the disk midplane.
As the settling process takes place on the disk,
dust particles will leave the upper layer
leaving behind a  
depleted small
grains population, having
a dust-to-gas mass ratio, $\zeta_{\rm small}$, lower than 
the initial value, $\zeta_{\rm std}$. On the other hand, 
the big grain population will have a dust-to-gas mass ratio, $\zeta_{\rm big}$, larger
than the initial value, since small dust particles
being settled from upper layers will add mass to this population. 
Therefore, a decrease of $\zeta_{\rm small}$
automatically implies an increase (of the same proportion) of $\zeta_{\rm big}$.
We parametrize settling with the parameter
$\epsilon$, referred to as the  dust-to-gas mass ratio and defined as
$\epsilon = \zeta_{\rm small}/\zeta_{\rm std}$, i.e., the mass
fraction of the small grains relative to the standard
value.
Therefore, lower values of $\epsilon$ represent more settled disks.

\subsection{Spectral indices}
\label{sec:spec_ind}
\indent 
Since evolutionary effects, like dust settling and grain growth, are more apparent
at longer wavelengths, we made 
plots of spectral indices using the PACS photometry reported here
and IRAC and MIPS fluxes from H07a,
in order
to determine the 
overall 
degree of dust settling in our disks.

\begin{figure*}
\centering
\includegraphics[width=0.7\paperwidth]{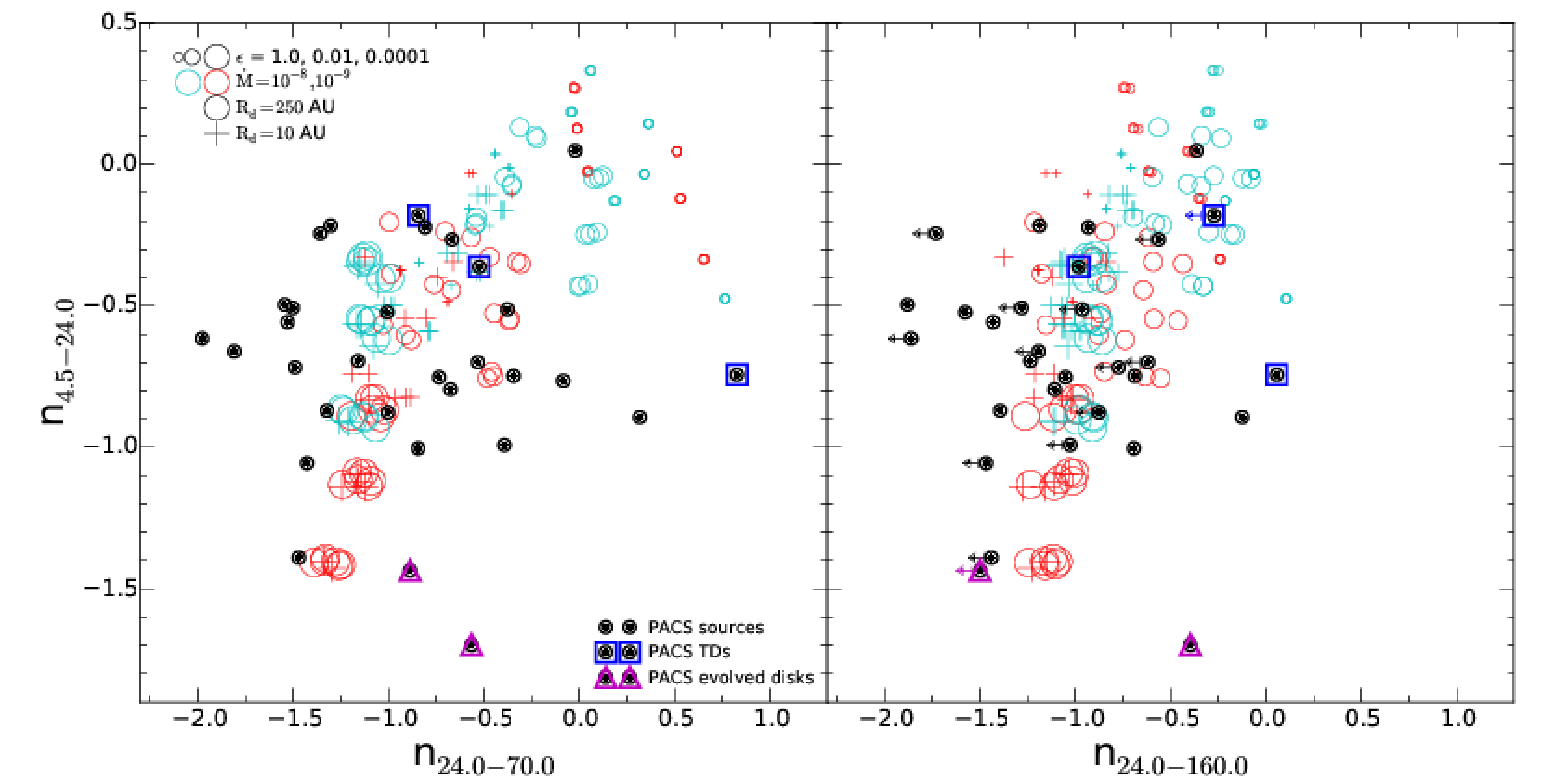}
\caption{$\rm n_{4.5-24}$ vs $\rm n_{24-70}$ (left) and $\rm n_{4.5-24}$ vs $\rm n_{24-160}$ (right)
for objects in the \SOri cluster 
(solid circles) and disk models (open circles: models with $\rm R_{\rm d}$ = 250 AU.
Plus signs: models with $\rm R_{\rm d}$ = 10 AU). 
TDs are surrounded by blue squares and evolved disks by magenta triangles. 
The models use a star of $\rm M_{*}=0.5$ $\rm M_{\odot}$,
$\rm R_{*}=2$ $\rm R_{\odot}$, $\rm T_{\rm eff}=4000$ K
and two values of mass accretion rates, $\rm \dot M = 10^{-8}$ $\rm M_{\odot}yr^{-1}$ (light blue symbols)
and  $\rm \dot M = 10^{-9}$ $\rm M_{\odot}yr^{-1}$ (red symbols).
Decreasing values of $\epsilon$ (more settled disks) are plotted with larger symbol sizes
($\epsilon$ = 1.0, 0.01 and 0.0001).
Error bars are included, but in most cases are smaller than the symbol size.
Arrows represent upper limits. 
Note the clear separation of the $\epsilon$ parameter on the x axis. The models are from 
D'Alessio et al. (2006). The \SOri IRAC and MIPS photometry is taken from H07a.}
\label{fig:nmod1} 
\end{figure*}

The left panel of Figure~\ref{fig:nmod1} displays the
$\rm n_{4.5-24}$ vs $\rm n_{24-70}$ diagram for
\SOri members (solid circles). 
We compare the observed spectral indices
with theoretical indices calculated for a
set of models with the following parameters:
$\rm M_{*}=0.5$ $\rm M_{\odot}$,
$\rm R_{*}=2$ $\rm R_{\odot}$, $\rm T_{\rm eff}=4000$ K
(all these parameters constitutes mean values in the cluster
as shown in Tables~\ref{tab:properties} and~\ref{tab:properties_undetec})
and two values of
mass accretion rates $\rm \dot M=$ $10^{-8}$ $\rm M_{\odot}yr^{-1}$ (light blue symbols) 
and  $\rm \dot M=$ $10^{-9}$ $\rm M_{\odot}yr^{-1}$ (red symbols),
where the latter is a common value for \SOri sources (Figure~\ref{fig:Mdot_histo}).
We fixed the viscosity $\alpha$ to a typical value of 0.01 \citep{hartmann98}. 
As we already said in section~\ref{sec:HR_D},
we used a distance to the \SOri
cluster of 385 pc.
Predicted spectral indices are indicated by open circles 
for models with $\rm R_{\rm d}$ = 250 AU,
and plus signs for models with $\rm R_{\rm d}$ = 10 AU, where the
size of the symbol varies according to the degree of dust settling,
with larger symbols representing
more settled disks (lower values of $\epsilon$).  
Squares and triangles account for
TD/PTD disks and evolved disks, respectively.
The scatter observed in each 
case represents the variety of properties like maximum grain size at
the disk midplane, $a_{\rm maxb}$, the maximum grain size
at the disk wall, $a_{\rm maxw}$, 
disk inclination angle, $\mu$, disk external radius, $\rm R_{\rm d}$, and ice abundances adopted in the models.
A similar plot is shown in the right panel of Figure~\ref{fig:nmod1}
for the spectral index $\rm n_{24-160}$ in the x axis. 
As shown, models with different degree of settling populate different
regions on this diagrams, being the more settled disks bluer than 
the less settled ones. 
Note as well the dependence in mass accretion rates on the $\rm n_{4.5-24}$ slope.
Models with $\rm \dot M$ = 10$^{-9}$ $\rm M_{\odot}yr^{-1}$ have $\rm n_{4.5-24}$ spectral indices as low as
$-1.5$, while models with $\rm \dot M$ = 10$^{-8}$ $\rm M_{\odot}yr^{-1}$ have indices higher than $-1.0$.
As can be inferred from these diagrams, the majority of our objects
falls in regions of $\epsilon$ $\leq$ 0.01, indicating that
most of the \SOri disks have significant degree of settling.
In both cases (left and right panels of Figure~\ref{fig:nmod1}) 
the indices $\rm n_{24-70}$ and $\rm n_{24-160}$ for a set of objects
lay outside 
the region populated by the models. However, 
in \S\ref{sec:modelling}
we modeled almost all
of these outliers.

\begin{figure*}
\centering
\includegraphics[width=0.4\paperwidth]{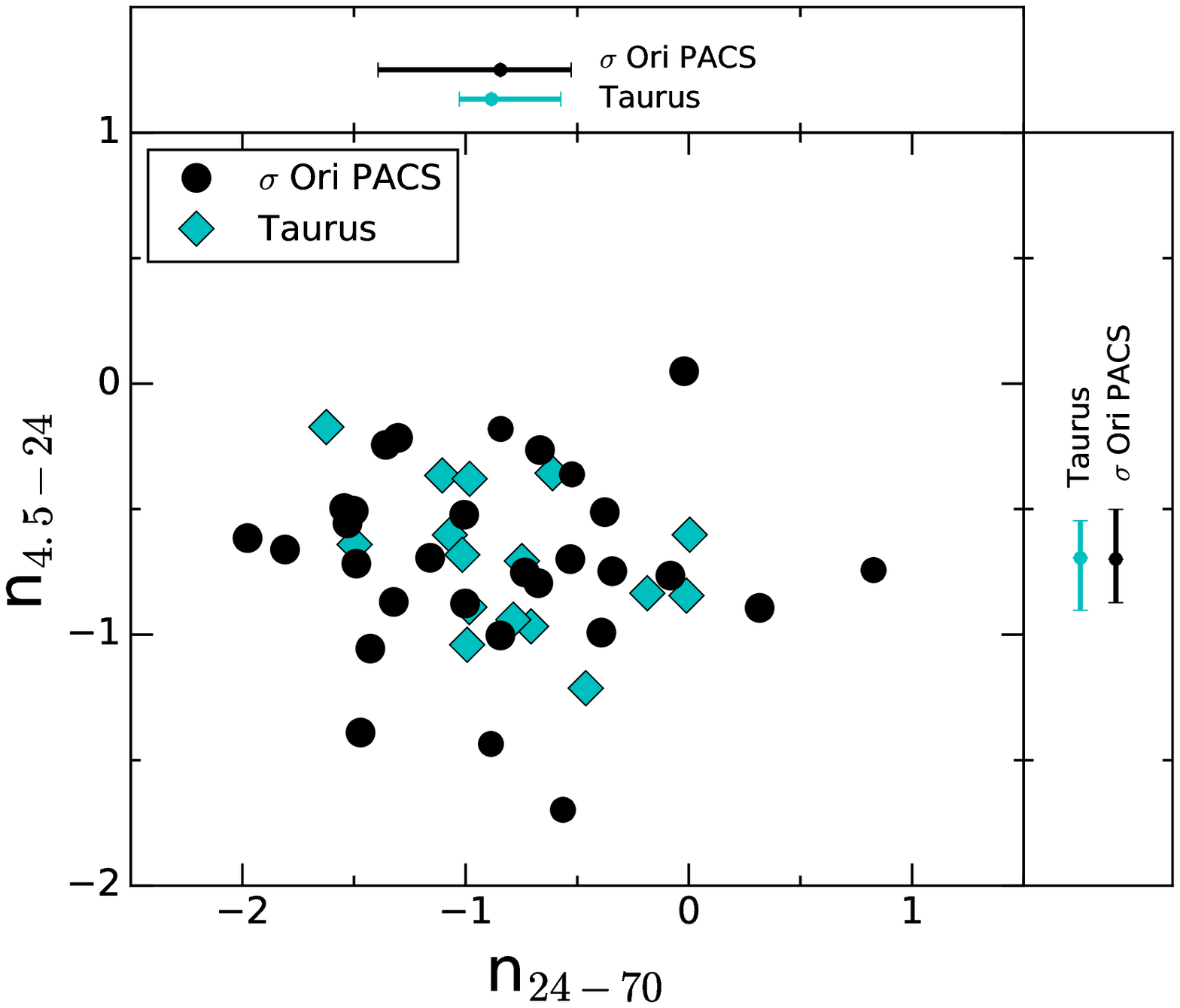}
\caption{$\rm n_{4.5-24}$ vs $\rm n_{24-70}$ for objects in the \SOri cluster 
(black dots). The Taurus star-forming
region (light blue diamonds) is plotted for comparison,
where the photometric data was taken from \citet{howard13,luhman10}. 
Also shown are the medians and quartiles for the $\rm n_{24-70}$ (top) and  
$\rm n_{4.5-24}$ (right) indices of each region (error bars). 
\SOri spectral indices are similar to Taurus.}
\label{fig:nobs1} 
\end{figure*}

Figure~\ref{fig:nobs1} displays the $\rm n_{4.5-24}$ vs $\rm n_{24-70}$
spectral indices of \SOri sources 
(black dots)
compared to disks in Taurus (1-3 Myr, diamonds) 
from \citet{howard13,luhman10}. 
Also shown are the medians and quartiles for the $\rm n_{24-70}$ (top) and  
$\rm n_{4.5-24}$ (right) indices of each region for comparison.
The \SOri sample has spectral indices similar to the population of Taurus.  
  
\begin{figure*}
\centering
\includegraphics[width=0.7\paperwidth]{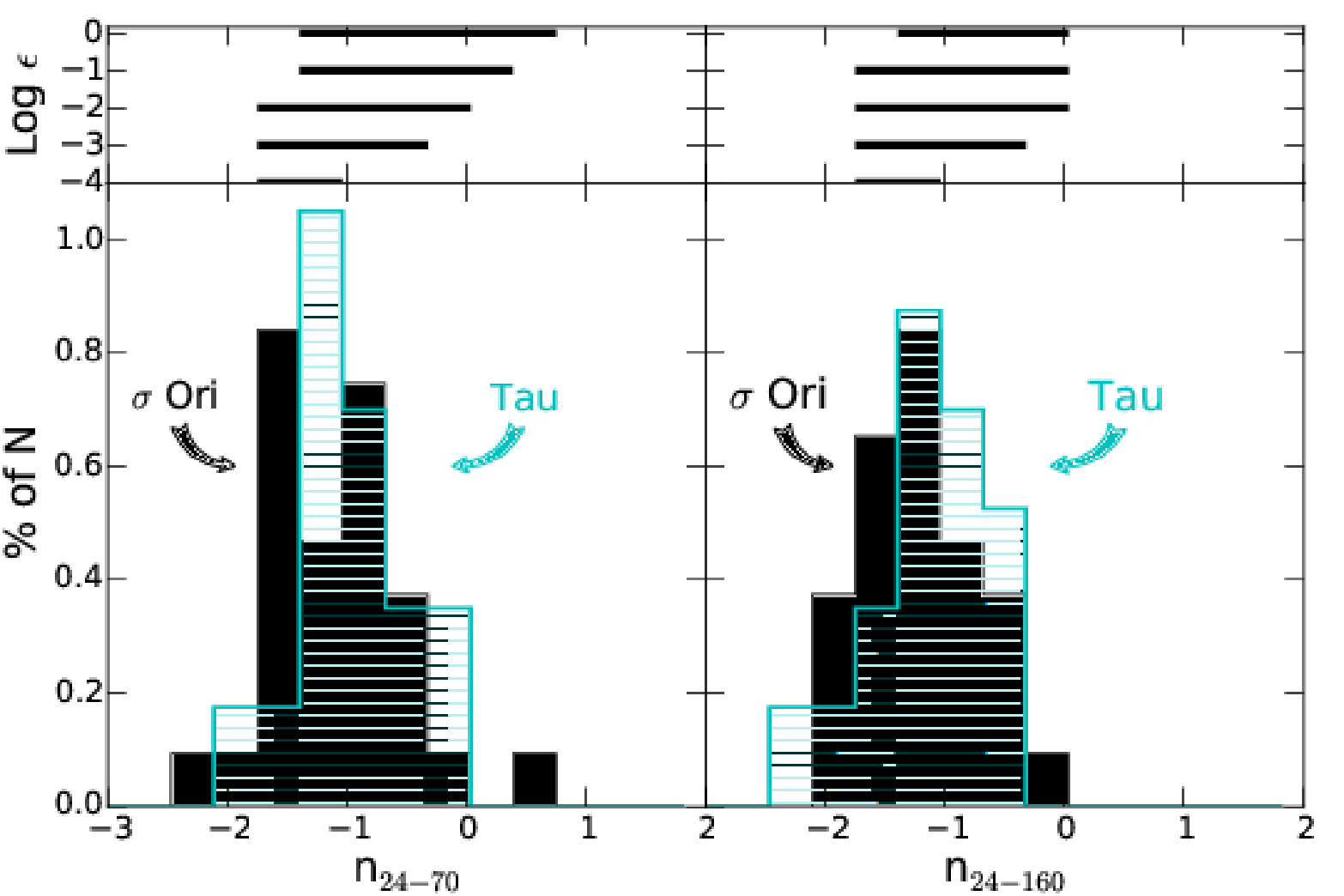}
\caption{Bottom: histograms of $\rm n_{24-70}$ (left) and 
$\rm n_{24-160}$ (right) for objects in the \SOri and Taurus regions.
Top: $\rm n_{24-70}$ (left) and $\rm n_{24-160}$ (right) for models 
with different degrees of settling: $\epsilon$ = 1.0, 0.1, 0.01, 0.001 and 0.0001 from top to bottom.
The models are from \citet{dalessio06}. 
The \SOri MIPS photometry is from H07a. The Taurus 
PACS fluxes were taken from \citet{howard13}, while the MIPS photometry 
was provided by \citet{luhman10}. Note how disks in \SOri have higher
degree of dust settling, since there is a significant number 
of PACS sources with lower values of $\epsilon$
compared to disks in Taurus.}
\label{fig:hist}	 
\end{figure*}

In order to inquire deeply on this subject
as well as understand the degree of dust
settling in the sample,
we compared in Figure~\ref{fig:hist} histograms of observed
$\rm n_{24-70}$ and $\rm n_{24-160}$ indices
for \SOri (black) and Taurus (light-blue)
disks.
The ranges of the spectral indices for the disks models shown on Figure~\ref{fig:nmod1}, 
were plotted as bands on 
top of these figures for $\epsilon$ values of
1.0, 0.1, 0.01, 0.001 and 0.0001 from top to bottom.
The Taurus histograms were made with 
MIPS 24 $\mu m$ photometry from \citet{luhman10} while the PACS 
fluxes were taken from \citet{howard13}, giving an 
overall average of 21 sources.
As shown on the left panel, 
most of our objects have considerable degree of dust settling, 
since the peak of the histogram, labeled as $\sigma$ Ori, 
falls at the same range as models with $\epsilon$ $\leq$ 0.01.
A similar situation happens for the $\rm n_{24-160}$ index,
shown on the right panel.
In this case, however, there is more overlap between ranges occupied by models with different
degrees of settling.
These results imply that disks in our 
sample have experienced some evolution over time.
Note as well, 
how disks in Taurus have lower 
degree of dust settling, since the peaks of the Taurus histograms fall slightly 
to the right (bigger values of $\epsilon$) compared to 
a significant number of \SOri sources with $\epsilon \leq 0.01$ in the figures. This is 
consistent with previous studies of dust evolution where 
older disks are more settled and emit less IR excesses than young ones, which causes 
a decrease in the disk fraction as function of age for a given stellar group (H07a).
It is important to note that, the fact that 
disks on Taurus already exhibit signs of dust 
settling indicates that this process must happen 
at an early evolutionary stage.

\subsection{Modeling of Individual Objects \label{sec:modelling}}

\indent
To better constrain the dust content in our PACS disks classified as class II stars 
and EV disks (the TDs will be studied in a future paper),
we modeled the SEDs of individual objects with irradiated accretion disks models,
using as input the stellar properties and accretion rates reported in Table~\ref{tab:properties},
assuming a viscosity parameter $\alpha=0.01$ and a distance to \SOri 
of 385 pc (\S~\ref{sec:HR_D}).
For each object we calculated a 
total of 1800 disk models varying
the outer radius ($\rm R_{\rm d}$),
the degree of dust settling ($\epsilon$), 
the cosine of the inclination angle ($\mu$), 
the maximum grain size at the disk midplane ($a_{\rm maxb}$)
and at the wall ($a_{\rm maxw}$). 
Table~\ref{tab:model_par} lists 
all the relevant parameters.
We assumed a maximum grain size at the disk upper layer of 
0.25 $\mu$m.
We selected as best fit the model that yielded the minimum
value of  $\chi^{2}$.
Upper limits in the observed fluxes where excluded from the calculation
of the $\chi^{2}$ value. 

\begin{deluxetable}{cc}
\tablewidth{0pt}
\tablecaption{Model Parameters \label{tab:model_par}}
\tablehead{
\colhead{Parameter} & \colhead{value}
\\
}
\startdata
$\rm R_{\rm d}$ (AU)\tablenotemark{a} ...................  & 10, 30, 50, 70, 100, 150, 200, 250, 300, 350\\
$\epsilon$ ................................ & 1, 0.1, 0.01, 0.001, 0.0001\\
$\mu$ ................................ & 0.3, 0.6, 0.9 \\
$a_{\rm maxb}$ .......................... & 100 $\mu m$, 1 mm, 10 cm\\
$a_{\rm maxw}$ ($\mu m$) ................ & 0.25, 10\\
\enddata
\tablenotetext{a}{The source SO697 needed a smaller $\rm R_{\rm d}$ (Table~\ref{tab:fit_par}).}

\end{deluxetable}

\indent
We modeled 18 sources
out of 27 in the sample considered here 
(SO540 is taken as a TD, see section~\ref{sec:cha_PACS}): 
17 class II stars, and one EV disk.
One of these objects, SO844, has mm detections taken with SCUBA-2 at 850 $\mu$m
and the rest, only have upper limits \citep{williams13}. 
Of the 9 stars not modelled in this work, the source SO984 also has
observations at 1300 $\mu$m taken with 
the SubMillimeter Array \citep{williams13}. 
This object exhibits a MIPS 24 $\mu$m flux
lower than the expected for class II stars (see Figure~\ref{fig:seds})
resembling that of PTDs, so it will be model, separately, in a future work.
The rest are objects with significant
uncertainties in their estimations of
extinction and spectral types
or objects without mass accretion rates (\S~\ref{sec:HR_D}), 
so they were not modeled here.

Figure~\ref{fig:modelling} shows
the SEDs of our PACS disks (open circles) with the resulting fit (solid lines).
Dashed lines indicate the \SOri
median (Table~\ref{tab:Median}). 
Photospheric fluxes, using the colors of \citet{kenyon95},
are indicated by dotted lines.
Table~\ref{tab:fit_par} gives the parameters of the best fit model for each 
object with its corresponding reduced $\chi^{2}$ value ($\rm \chi^{2}_{red}$)
in the following order: target name, disk outer radius ($\rm R_{d}$),
disk outer radius confidence interval ($\rm R_{d}^{-} - R_{d}^{+}$), 
degree of dust settling ($\epsilon$), cosine of the inclination angle ($\mu$),
maximum grain size at the disk midplane ($a_{\rm maxb}$),
maximum grain size at the wall ($a_{\rm maxw}$), disk mass ($\rm M_{d}$)
and disk mass confidence interval ($\rm M_{d}^{-}-M_{d}^{+}$).

\begin{figure*}
\centering
\includegraphics[width=0.7\paperwidth]{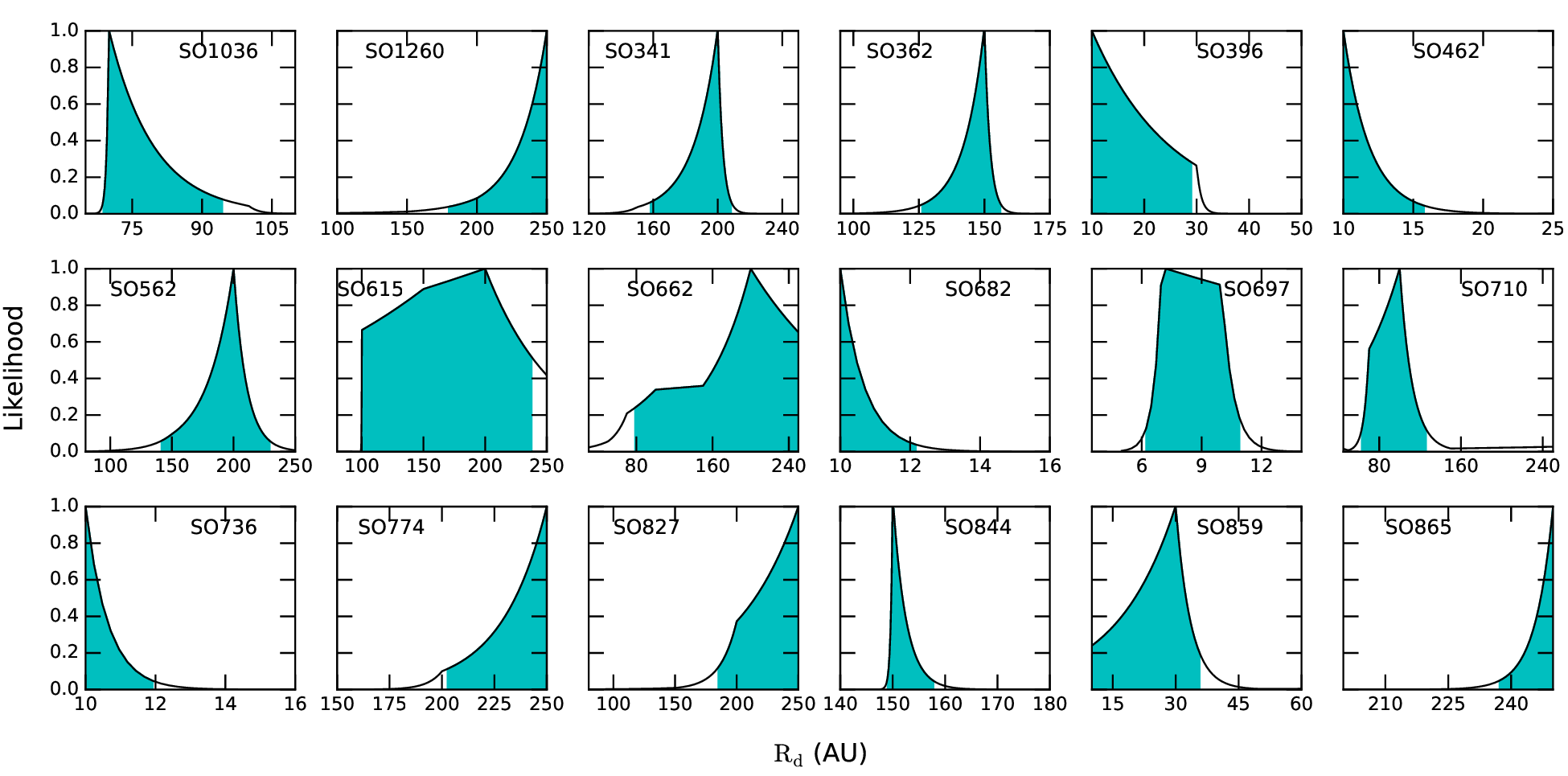}
\caption{Likelihood function, ${\cal L}$, {\it versus} disk radius, $\rm R_{d}$, for each source. 
The $x$ axis has been restrained around the maximum peak in each case 
in order to better visualize the likelihood function.
Confidence intervals of $\rm R_{d}$ are shown as light-blue shaded regions 
and are defined as the intervals that enclose 
95\%\ of the total area of ${\cal L}$.}
\label{fig:chi2}	 
\end{figure*}

In order to calculate confidence intervals of the outer radius of the disks, we first
estimated the likelihood function, ${\cal L}$, which is related to the
$\chi^{2}$ values through the expression ${\cal L} = \rm exp(-\chi^{2}/2)$. 
Since $\chi^2$ is a multidimensional function, at every radius we have several 
values of $\chi^2$, one for each one of the calculated models. Thus, the likelihood 
${\cal L}$ is computed using the minimum $\chi^{2}$ value at each radius. In 
Figure~\ref{fig:chi2} we show plots of the normalized likelihood as a function of radius 
for all the 18 sources that were successfully modeled.
We have restrained the $x$ axis around the maximum peak in each case 
in order to better visualize the likelihood function.
As usual, the confidence intervals of $\rm R_{d}$ are given as those radii 
$\rm R_{d,1}$ and $\rm R_{d,2}$ at which the area below the likelihood curve is 95\%\ of its 
total area \citep{sivia12}. These intervals are indicated by
light-blue shaded regions in each panel 
and are reported in column 3 
of Table~\ref{tab:fit_par}. 
For those cases where the best radius falls on one of the edges of the
range of radii used in the models, we have considered these values as upper or lower
limits, and are indicated by parenthesis instead of square brackets in Table~\ref{tab:fit_par}.
On the other hand, since the mass of the disk 
is a parameter that correlates with size, the confidence intervals of the 
mass are given as the values of the mass given by the best fit model at $\rm R_{d,1}$ and $\rm R_{d,2}$. 
These values are reported 
in column 9 of Table~\ref{tab:fit_par}. 

Some sources have peculiar SEDs: 
SO844 shows strong excess emission from the mid-IR to the far-IR, 
particularly in all four IRAC bands. This 
is probably due to variability. 
SO462, on the other hand, has a poor fit from H to IRAC 5.8 $\mu$m,
which significantly increases the value of $\chi^{2}$;
it is worth noting that this source has the largest $A_{\rm v}$ (=2.94, see Table~\ref{tab:properties}).
Finally, SO697 
needed a smaller radius than the other sources ($\rm R_d$ = 7 AU). This
is due to its low flux at 70 $\mu$m combined with one of the lowest
160 $\mu$m upper limits. Therefore, any disk model with an outer radius greater than 7 AU
will overestimate the flux at 70 $\mu$m and will produce a flux greater than the upper limit at 160 $\mu$m.

We point out, that we used the values of $\chi^{2}$ as a guide in order to obtain
the model that provides the best fit to the photometry
of each source. However, the values of $\chi^{2}$ are expected
to be large. One of the most problematic points to
fit is the IRAC 8 $\mu$m band. In a real spectrum, this 
band contains a contribution of the 10 $\mu$m silicate
feature. The total flux of this feature is highly
dependent on the adopted composition of the silicates,
as detailed modeling of IRS spectra have shown
(cf. McClure et al. 2012, 2013a).
However, we felt that one photometric point did not give 
enough constraints to justify having the silicate composition
as an additional parameter. Another reason for the large
values of $\chi^{2}$ is that we are only including the
errors in the photometry, which are small, and do
not include other sources of uncertainty such as the
inherent uncertainties in the distance, luminosity,
spectral types, and mass accretion rates. 
In a forthcoming paper we will model these sources 
including their IRS spectra, in order to characterize their silicate 
features, which will improve the current fit.
Through this study we will characterize the inner parts of the disk 
in the sense of grain grow, dust processing and types
of silicates present in the disks.

\begin{figure*}
\centering
\includegraphics[width=0.7\paperwidth]{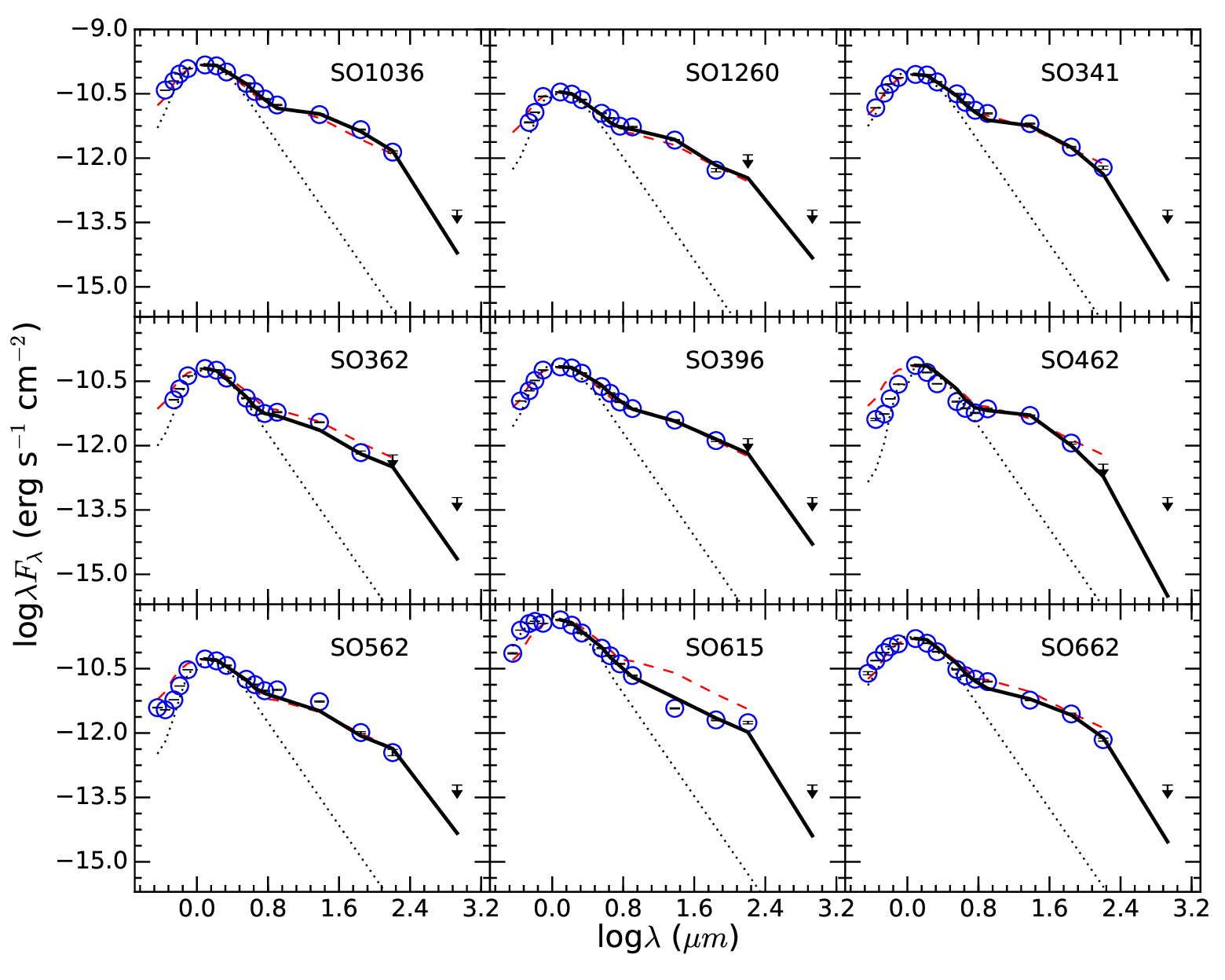}
\caption{Spectral energy distributions for \SOri sources including the mm photometry from \citet{williams13}. 
As in Figure~\ref{fig:seds}, photometry has
been dereddened following the Mathis law \citep[][$R$ = 3.1, open circles]{mathis90}.
Solid lines indicate the best fit model.
Dashed lines represent the median of \SOri PACS sources (Table~\ref{tab:Median}).
Dotted lines correspond to the photosphere-like fluxes 
using the colors of \citet{kenyon95}. Error bars are included, but
in most cases are smaller than the markersize. Downward arrows indicate upper limits.}
\label{fig:modelling} 
\end{figure*}

\begin{figure*}
\ContinuedFloat
\centering
\includegraphics[width=0.7\paperwidth]{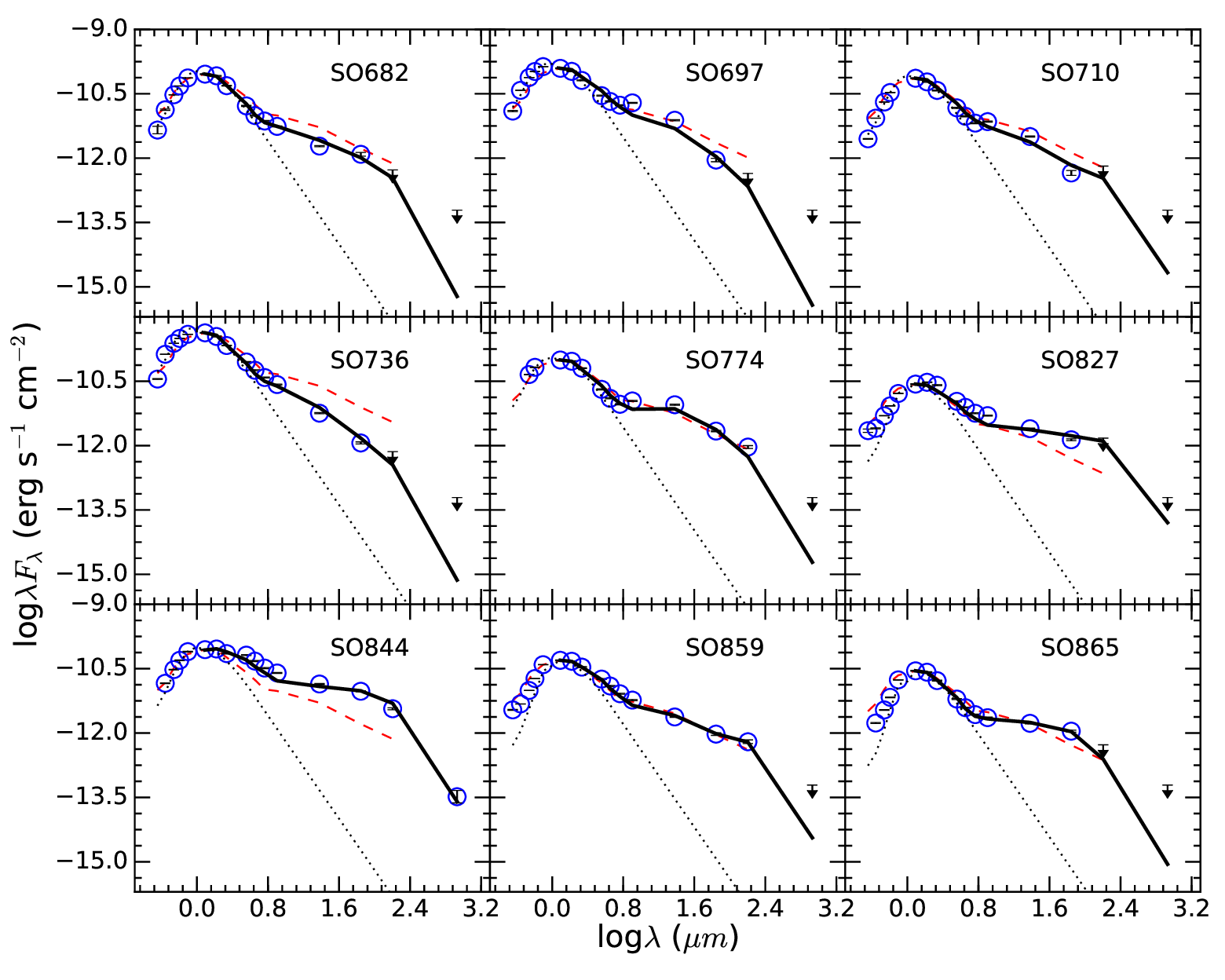}
\caption[]{caption (continued)}
\end{figure*}

\begin{deluxetable}{lrcrrrrrcr}
\centering
\tablewidth{0pt}
\tablecaption{Estimated Disk Parameters \label{tab:fit_par}}
\tablehead{
\multicolumn{1}{c}{Target} & \multicolumn{1}{c}{$\rm R_{\rm d}$} & \multicolumn{1}{c}{$\rm R_{\rm d}^{-}$ - $\rm R_{\rm d}^{+}$} & \multicolumn{1}{c}{$\epsilon$} & \multicolumn{1}{c}{$\mu$}& \multicolumn{1}{c}{$\rm a_{\rm maxb}$} & \multicolumn{1}{c}{$\rm a_{\rm maxw}$} & \multicolumn{1}{c}{$\rm M_{\rm d}$} &  \multicolumn{1}{c}{$\rm M_{\rm d}^{-}$ - $\rm M_{\rm d}^{+}$} & \multicolumn{1}{c}{$\rm \chi_{red}^{2}$}\\ \noalign{\smallskip}
\colhead{} & \colhead{(AU)} & \colhead{(AU)} & \colhead{} & \colhead{} & \colhead{} & \colhead{($\mu m$)} & \colhead{($10^{-3}\rm M_{\odot}$)} & \colhead{($10^{-3}\rm M_{\odot}$)} & \colhead{}
}
\startdata
SO1036 & 70  & [69 - 94] & 0.01   & 0.3 & 10 cm       & 10   & 5.19 & [4.78 - 6.10] & 6.116\\
SO1260 & 250 &[180 - 250) & 0.0001 & 0.9 & 100 $\mu m$ & 10   & 3.11 & [2.57 - 3.11) & 14.33\\
SO341 &  200 &[158 - 208] & 0.01   & 0.6 & 10 cm       & 10   & 1.24 & [1.05 - 1.34] & 24.784\\
SO362 &  150 &[126 - 156] & 0.0001 & 0.9 & 100 $\mu m$ & 0.25 & 1.01 & [0.91 - 1.11] & 31.489\\
SO396 &  10  & (10 - 29] &  0.01   & 0.3 & 100 $\mu m$ & 10   & 1.15 & (1.15 - 2.46] & 2.040\\
SO462 &  10  & (10 - 16] & 0.01   & 0.9 & 10 cm & 0.25 & 0.098 & (0.09 - 0.14] & 265.259\\
SO562 &  200 &[142 - 229] & 0.0001 & 0.9 & 100 $\mu m$ & 10   & 2.70 & [2.24- 3.26] & 55.817\\
SO615 &  200 &[100 - 238] & 0.0001 & 0.9 & 100 $\mu m$ & 10   & 1.25 & [0.68 - 1.49] & 51.825\\
SO662 &  200 &[79 - 250] & 0.001  & 0.9 & 10 cm       & 10   & 2.44 & [1.32 -2.91] & 52.388\\
SO682 &  10  &(10 - 12] & 0.01   & 0.6 & 1 mm        & 0.25 & 0.031 & (0.028 - 0.035] & 14.469\\
SO697 &  7   & [6 - 11] & 0.01   & 0.6 & 10 cm       & 10   & 0.11 & [0.10 - 0.16] & 110.766\\
SO710 &  100 & [63 - 126] & 0.0001 & 0.9 & 100 $\mu m$ & 10   & 0.69 & [0.45 - 0.85] & 44.761\\
SO736 &  10 & (10 - 12]  & 0.001  & 0.3 & 10 cm       & 0.25 & 0.22 & (0.22 - 0.27] & 27.975\\
SO774 &  250 & [203 - 250) & 0.01   & 0.9 & 10 cm       & 10   & 1.68 & [1.43 - 1.68) & 47.192\\
SO827 &  250 & [185 - 250) & 0.01   & 0.6 & 100 $\mu m$ & 10   & 4.94 & [3.90 - 4.60) & 34.874\\
SO844\tablenotemark{a} &  150 & [149 - 158] & 0.1    & 0.6 & 10 cm  & 10  & 39.40  & [39.40 - 43.70] & 48.159\\
SO859 &  30  & [10 - 36] & 0.01   & 0.3 & 100 $\mu m$ & 10   & 0.47 & [0.23 - 0.57] & 16.131\\
SO865 &  250 & [237 - 250) & 0.01   & 0.9 & 10 cm  & 0.25 & 0.85 & [0.77 - 0.85) & 3.450\\
\enddata
\tablenotetext{a}{Object with SCUBA-2 850 $\mu$m photometry from \citet{williams13}.}
\tablecomments{Column 1: ID following H07a; Column 2: disk outer radius; Column 3:
disk outer radius confidence interval;
Column 4: degree of
dust settling; Column 5: cosine of the inclination angle; Column 6: maximum grain 
size at the disk midplane; Column 7: maximum grain 
size at the disk wall; Column 8: disk mass; Column 9: disk mass confidence interval;
Column 10: $\rm \chi_{red}^{2}$ value of the fit. 
}
\end{deluxetable}	

The flux in the mid IR is mostly dictated by
the mass accretion rate and the inclination,
since for settled disks a large fraction of the
flux arises in the innermost disk regions. For instance,
80\% the flux at 24 $\mu$m and shorter wavelengths arises inside 
$\sim$2 AU for a disk with $\epsilon$ = 0.001,
instead of $\sim$20 AU for a well mixed disk \citep{dalessio06, mcclure13b}.
In addition, as settling increases the depleted upper layers
become optically thin, exposing the disk midplane.
Therefore, highly settled disks 
are colder and radiate less than disks
with low degree of settling \citep{dalessio06}.
We found that most of our objects (60\%) 
can be explained by
a significant degree of dust settling, $\epsilon = 0.01$, 
consistent with studies in young regions 
like Taurus and Ophiuchus \citep{furlan06,mcclure10}. 
We also found that some of our objects
have an even higher degree of dust settling ($\epsilon = 0.0001$),
which, in turn, is consistent with previous studies 
of disk frequency as a function of age 
in different young stellar populations (H07a).

\begin{figure*}
\centering
\centering
\includegraphics[width=0.7\paperwidth]{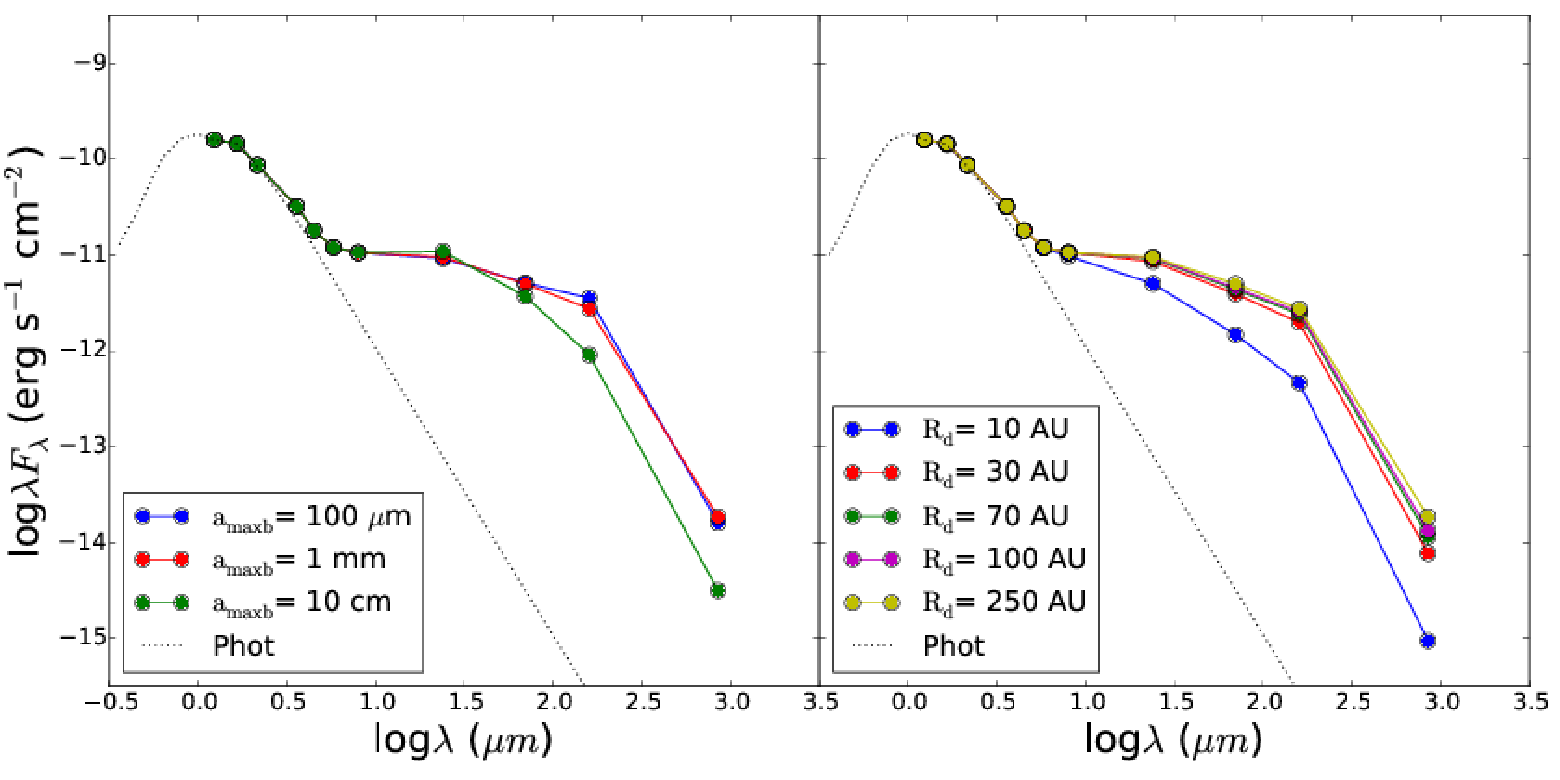}
\caption{
{\it Left}: theoretical SEDs with different values of $\rm a_{maxb}$.
Note how disks with $\rm a_{maxb}$ = 10 cm have far-IR fluxes substantially 
lower (almost an order of magnitude) than disks with smaller $\rm a_{maxb}$.
However, changes on the SED for $\rm a_{maxb}$ $\leq$ 1 mm
are less than 10\%, so we cannot discriminate them.
{\it Right}: theoretical SEDs with different sizes. 
Note how small disks with $\rm R_d$ = 10 AU exhibit lower
fluxes in the mid and far-IR compared to disks with larger sizes, reaching almost an order
of magnitude at 160 $\mu$m.
For the cases with  $\rm R_d$ $\geq$ 70 AU changes
between the SEDs at far-IR wavelengths are less than 10\%,
so we cannot discriminate them.
Dotted lines correspond to the photosphere-like fluxes 
for one of the stars in our sample
using the colors of \citet{kenyon95}.
For these models we used an $\epsilon$ value of 0.01. }
\label{fig:Rd_amaxb_ch} 
\end{figure*}

The dependence of the flux with $a_{\rm maxb}$
has been discussed by \citet{dalessio01}.
Even though we found 9 sources to have $\rm a_{maxb}$ = 100 $\mu$m (Table~\ref{tab:fit_par}),
the left panel of Figure~\ref{fig:Rd_amaxb_ch} shows that models with 
$\epsilon$ = 0.01 and $\rm a_{maxb}$ $\leq$ 1mm only differ by less than 10\%,
factor that is even smaller for $\epsilon$ = 0.0001,
so we cannot discriminate sizes of $a_{\rm maxb} \leq $ 1 mm.
The right panel, on the other hand, shows the effects of changing
the disk radius on the SED. As expected, smaller disks produced 
steeper mid and far-IR slopes and lower fluxes. 
Differences in disks radii are more apparent 
for $\rm R_d < 70$ AU.
We found that 61\% of our objects can be
modeled with large sizes, $\rm R_{\rm d} \geq 100$ AU.
The rest, have dust disk radii less than 80 AU.
Note how changes in $\rm R_d$ affects
the SED from MIPS 24 $\mu$m to mm wavelengths 
while changes in $\rm a_{maxb}$ are only apparent beyond 
24 $\mu$m.  

We found no correlation between the mass accretion rate
and the degree of dust settling. The disks with highest and lowest 
mass accretion rate in our sample have $\epsilon = 0.01$. This
seems to imply that the turbulence expected in the 
high accretors is not influencing the vertical
distribution of dust in disks, which is contrary to
expectations. 

\section{Discussion: Disk evolution in the \SOri cluster \label{sec:disscusion}}

\subsection{Disk properties}

We have obtained detailed disk structures
for 18 disks in the \SOri cluster using irradiated
accretion disk models \citep{dalessio06}, 
constrained by
mass accretion rates estimated independently 
from optical spectra.
As a result of our modeling, we 
inferred disk masses and radii,
assuming a
gas-to-dust ratio close to that of the ISM. 
The masses of the best fit models are given in 
Table~\ref{tab:fit_par}; most of the
masses are 
of the order of
$\sim$$10^{-3}$ to 10$^{-4}$ M$_{\odot}$.
By analyzing possible degeneracies in SED fitting
using a set of 1800 generic models 
with different parameters, we estimated that 
the uncertainty in 
the inferred 
disk masses 
is a factor of 3,
if the mass accretion rate of the sources are known.

Our mass determinations are consistent with those
inferred 
by \citet{williams13}
from 850 $\mu$m SCUBA-2 observations.
\citet{williams13} detected 9 disks, but
only
five of these
lie inside the FOV of our 
PACS observations. 
Four of these sources (SO540, SO844, SO984 and SO1153) 
were detected at 70 and 160 $\mu$m.
The only source not detected, SO609, is a class III
star (H07a). 
Of the detections, SO1153 is a class I star, as seen in Figure~\ref{fig:seds},
and it was not modeled here.
Similarly, the source 
SO984 will be modeled in a future work
together with other TDs
such as SO540 (see section~\ref{sec:modelling}). 
The SED and best-fit model for the source SO844
is shown in Figure~\ref{fig:modelling}
with inferred properties in Table~\ref{tab:fit_par}.
The disk mass for this object,
39 M$_{\rm Jup}$, is a 
factor of $\sim$8 higher than the mass estimated by \citet{williams13}.
The reason for this is that our
dust grain opacity ($\kappa_{\nu}$) at 850$\mu$m for this object 
is significantly lower
than them.
Unlike \citet{williams13}, which assumed a dust opacity 
of $\rm \kappa_{\nu} = 0.1(\nu/1200 GHz)$ $\rm cm^{2}g^{-1}$,  
we used a consistent
opacity law (estimated through detailed modeling of the SEDs) 
for each one of our objects. This opacity depends on the mix of materials assumed 
in our disk models (silicates, graphite, water, etc), their abundances and
their grain size distributions.
Therefore, each object has a distinct disk dust opacity.
With our dust opacity we can reproduce not only the flux
at 850 $\mu$m, but the entire SED (cf. Figure~\ref{fig:modelling}).

\begin{figure}[ht]
\includegraphics[width=1\columnwidth]{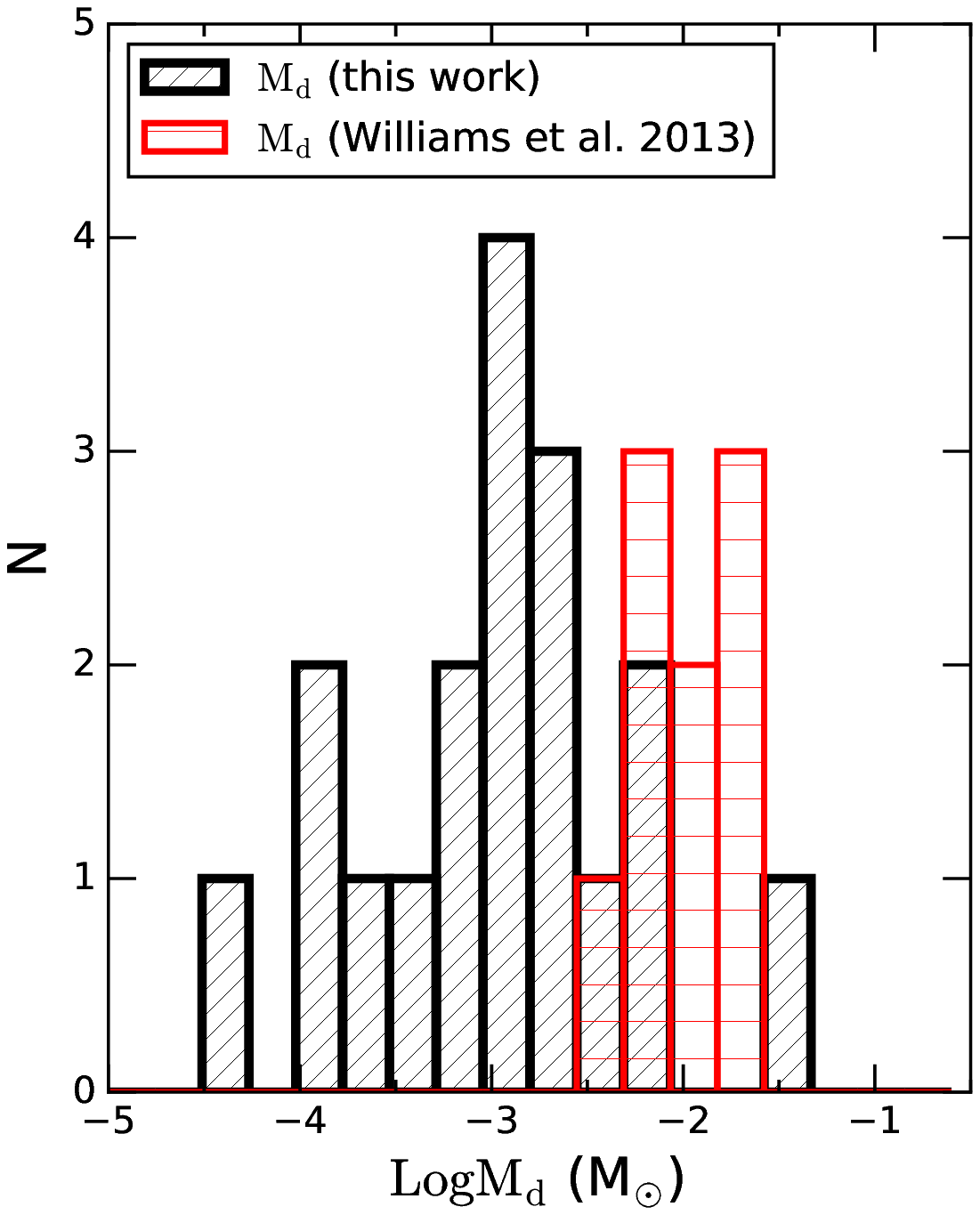}
\caption{Disk mass ($\rm M_{d}$) distribution for \SOri members with PACS detections reported
on Table~\ref{tab:fit_par} (black). 
The mass distribution found by \citet{williams13} is shown in red for comparison.
All our disks, but SO844 ($\rm M_{d}$ = 39 $\rm M_{Jup}$), correspond to the lower end of the cluster 
mass distribution while the William's disks are located
at the high mass end. This plot shows how PACS observations can be more suitable in detecting less massive 
small disks than mm observations.
}
\label{fig:hist_Mds_KW}	 
\end{figure}

Figure~\ref{fig:hist_Mds_KW} shows the disk mass
distribution for our \SOri sources in black 
compared to the mass distribution found by \citet{williams13} in red.  
Our detections include significantly lower values than the 
\citet{williams13} survey. 
The fact that
we have detected 24 more sources than \citet{williams13} inside the FOV of their observations,
indicates that the
PACS photometry was more suitable 
for detecting even small young disks than the shallow submillimeter observations.
However, since
our longest wavelength is
160$\mu$m, 
we may be missing emission
from the largest grains and underestimating 
our masses. 
Our models are consistent with the upper limits of 
the 850 $\mu$m SCUBA-2 observations (Figure~\ref{fig:modelling}), which 
suggests that the mass deficit is not large.
Moreover, \citet{williams13} reported a 3$\sigma$ limit of $\rm 4.3x10^{-3}$ $\rm M_{\odot} \sim 4.5$ $\rm M_{Jup}$, 
so their non detections
imply disks masses $<$ 5 $\rm M_{Jup}$, 
a value that is very similar to the upper limit of our disks masses for sources not detected with mm observations 
($\sim$5.5 $\rm M_{Jup}$, source SO1036. See Table 7).
This supports the fact that we are truly looking at disks with very low masses.
More sensitive
mm observations are still required to determine this mass deficit.
In any event, the masses of the \SOri disks besides SO844, range between 0.03 to
$\sim$6 M$_{\rm Jup}$, 
with 35\% 
lower than 1 $\rm M_{Jup}$.
Like \citet{williams13},
we conclude that Jupiter scale giant planet formation must be 
complete in these objects, 
which indicates that either giant planets form in 
less than 3 Myr or that it is difficult to
make them in clustered environments. 

\begin{figure}[ht]
\includegraphics[width=1\columnwidth]{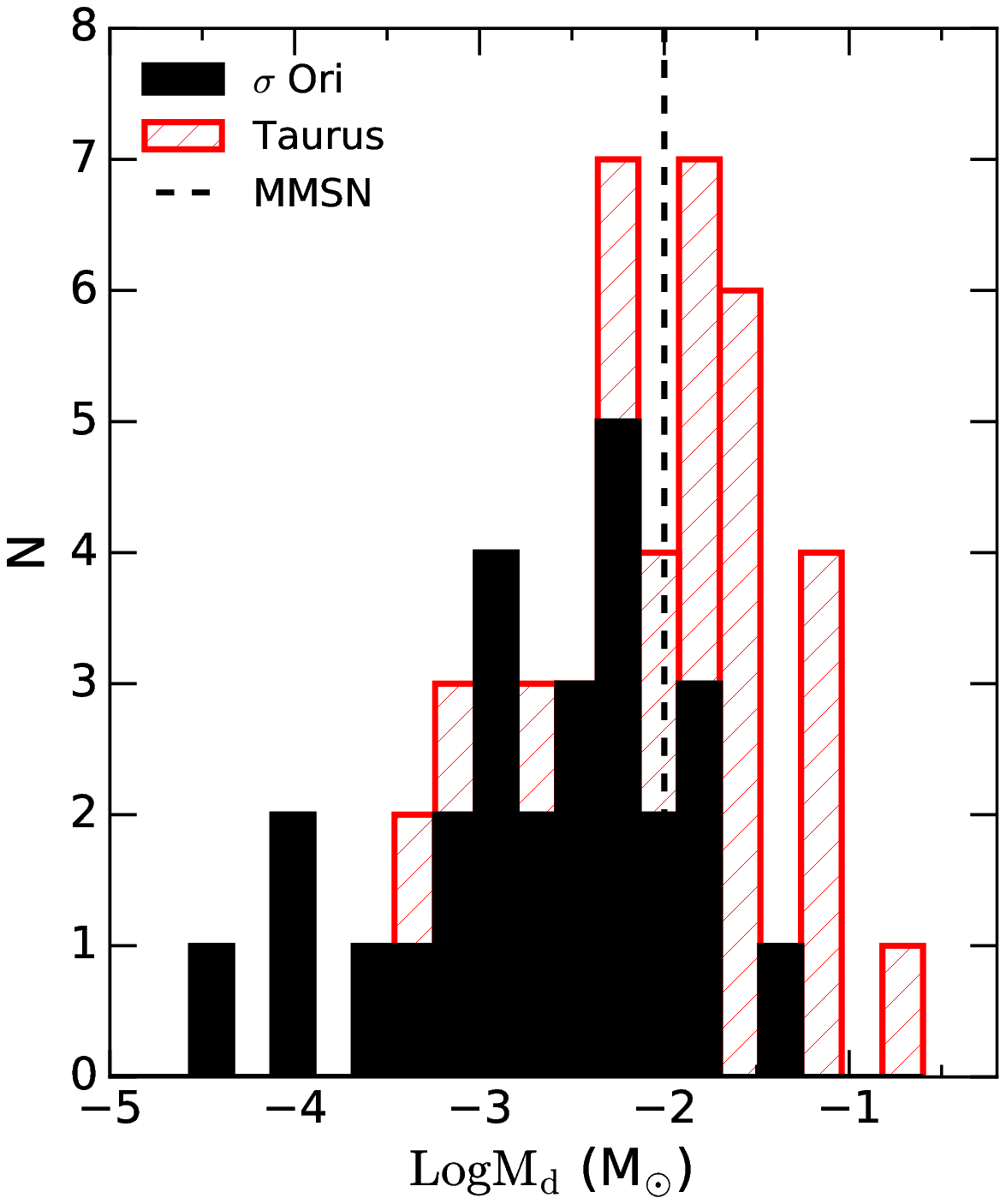}
\caption{Disk mass ($\rm M_{d}$) histograms for disks in \SOri (black)
and Taurus members (red) reported in \citet{andrews05}.
The disk mass distribution for \SOri disks includes our sample 
(Table~\ref{tab:fit_par}) and sources reported by \citet{williams13}.
Also shown is the MMSN of 10 $\rm M_{Jup}$ (dashed line). 
\SOri disk masses range
between 0.03 and $\sim$39 $\rm M_{Jup}$. Most 
objects have masses significantly lower than disk masses in Taurus.
}
\label{fig:hist_Mds_tau}	 
\end{figure}

Figure~\ref{fig:hist_Mds_tau} 
shows the disk mass distribution of \SOri (our sample and Williams') compared to that
of Taurus. An overall decrease in disk mass is clearly seen. 
This behavior of disk mass with age is expected from
viscous evolution (Hartmann et al. 1998). For instance, 
between 1-3 Myr, the ages of Taurus and
$\sigma$ Ori, the disk mass should decrease by a factor
of 1.5. However, the disk sizes should correspondingly 
increase by a factor of 2.5, which we do not see (see next section and Table~\ref{tab:fit_par}).
One possibility is that these disks have been subject to radial drift,
according to which the millimeter-size grains have drifted 
inwards and have piled up into pressure bumps, 
resulting in an observationally smaller dust disk \citep{birnstiel14,pinilla12b}
an effect that may be at play in TW Hya \citep{andrews12} and other
disks \citep{panic09,pinilla12a,laibe12,rosenfeld13,zhang14}. 
Another possible explanation for the presence of small disks is due to
interactions with stellar companions.
This effect 
has been observed in
Taurus for close binaries
with separations $<$40 AU \citep{kraus12}.
Multiplicity in \SOri has been reviewed by \citet{caballero14}, where he
notes that
outside the central arcminute, only ten close binaries have been
reported with angular separations between 0\arcsec.4-3.\arcsec0 ($\sim$150-1200 AU).
Of these, only two have been imaged with adaptive optics. 
In short, more high-resolution imaging surveys of close
binaries in \SOri are needed in order to determine if
multiplicity is a main factor affecting the evolution
of disks.  
Alternatively, these disks
may have suffered the effects of photoevaporation, in which
case both the dust and the gas have dissipated from the outer disk.
We explore this possibility in the next section.

\subsection{Disk photoevaporation in \SOri}

The age of the \SOri cluster is estimated to be
$\sim$3 Myr (H07a), older than Taurus \citep[][$\sim$1-2 Myr]{kenyon95} 
or the ONC \citep[][$<$1 Myr]{hillenbrand97}.
Therefore, \SOri disks must have evolved relative to these
younger clusters. 
By applying Kolmogorov-Smirnov (K-S) tests 
for the $\rm n_{24-70}$ 
and $\rm n_{24-160}$ spectral indices 
between disks in Taurus and $\sigma$ Ori (see Figure~\ref{fig:hist}),
we found that the levels of significance $p$
for these indices are not small enough (compared to the statistics $D$) as to say that the 
distributions are intrinsically different (Table~\ref{tab:KS_test}).
However, testing differences in the inner parts of 
the disks between both distributions (using $\rm n_{\rm K-24}$, $\rm n_{\rm K-70}$ and 
$\rm n_{\rm K-160}$ spectral indices), shows that, as the wavelength 
increases, the distributions become substantially different, such that 
for the case of the last index, $\rm n_{\rm K-160}$, 
we can reject the null hypothesis that the samples are drawn from the same distribution
($D > p$, Table~\ref{tab:KS_test}). 
This may imply that
dissipation due to nearby massive stars
dominates in the \SOri cluster,
photoevaporating the outer disks, in contrast to the inside-out 
dissipation in regions without massive stars, like Taurus.

Since disks in \SOri are accretion disks, with substantial mass accretion rates
(\S~\ref{sec:HR_D}), viscous evolution must also be at play. 
Two alternatives 
are then possible, isolated viscous evolution or viscous
evolution mediated by external photoevaporation. 
To consider photoevaporation-mediated viscous evolution as a
possibility, we must first consider its feasibility. 
To do so, we estimated the field intensity, $\rm G_{0}$, 
produced by 
the three brightest objects in the system, the $\sigma$ Ori Aa,Ab 
(O9.5, B0.5) pair
and $\sigma$ Ori B 
($\sim$B0-B2) with a total mass of $\sim$ 44 $\rm M_{\odot}$ \citep{simondiaz15,caballero14}.
Using the reported stellar luminosities
for these stars, and assuming
that most of the stellar radiation is in the form of FUV photons,
we estimated the intensity of the incident  
radiation at a given distance $r$ from the source:

\begin{equation}
G_{0} = \frac{1}{F_{0}}\frac{L_{\rm FUV}}{4\pi r^{2}},
\end{equation}

\noindent
where $F_{0}$ is the typical interstellar flux level with a value
of 1.6 x 10$^{-3}$ erg s$^{-1}$ cm$^{-2}$
\citep{habing68}, 
$L_{FUV}$ is the FUV luminosity 
of the \SOri multiple system and 
$r$ the true distance to the ionizing sources. 
We used Montecarlo simulations to calculate the distribution of true distances
to the center for a given impact parameter, and from this the expected
distribution of $G_0$, following \citet{anderson13}. 

\begin{figure}[ht]
\includegraphics[width=1\columnwidth]{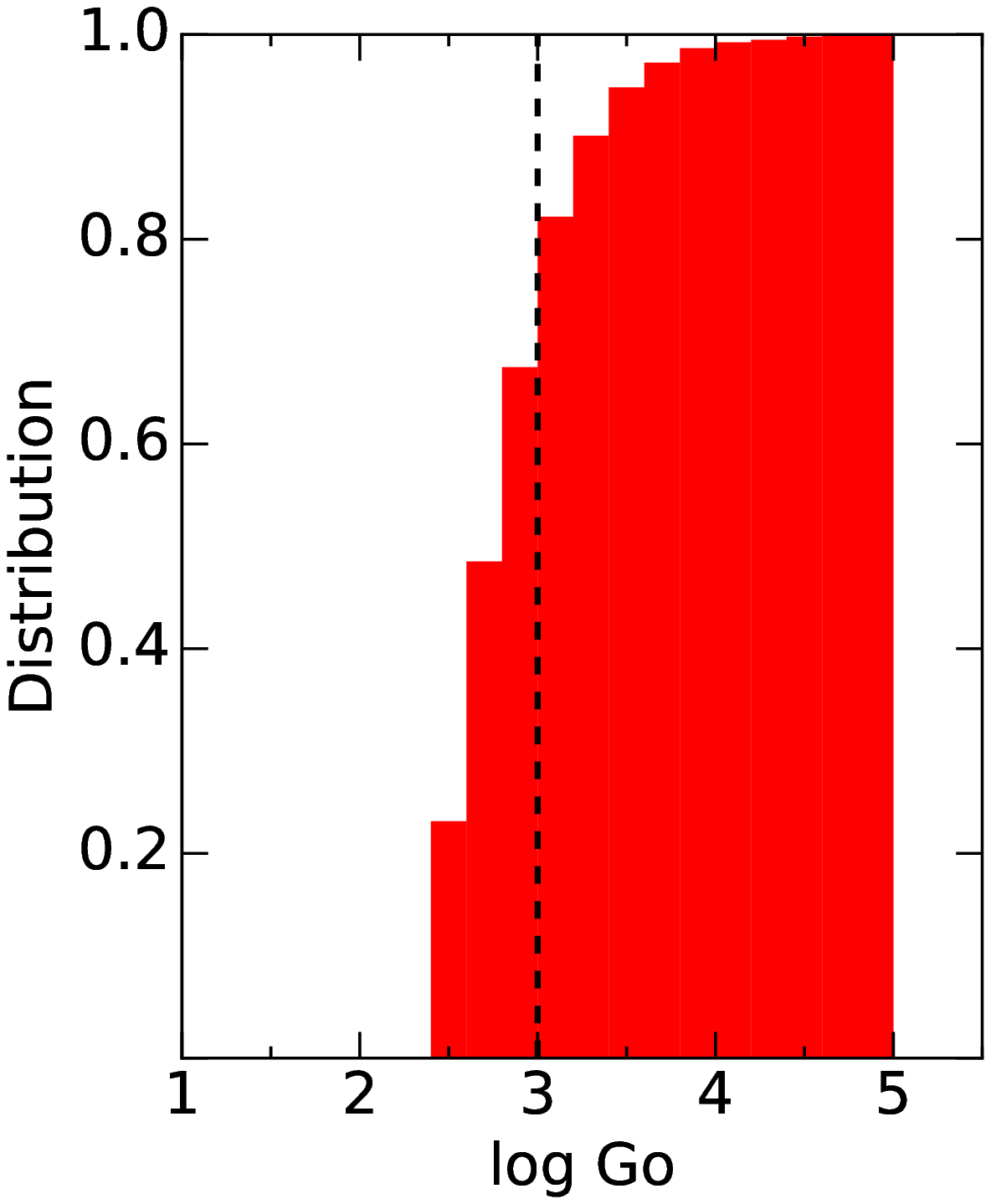}
\caption{Cumulative distribution of expected FUV fluxes for a sample of 32 PACS
disks in the \SOri cluster. Nearly 80\% of PACS disks are exposed to
flux levels of $\rm G_{0} \lesssim 1000$ (dashed line).}
\label{fig:FUV_hist} 
\end{figure}

\indent
We obtained that 80\% of our disks are exposed to flux levels
of $300 \lesssim \rm G_{0} \lesssim 1000$ (see Figure~\ref{fig:FUV_hist}). 
Even though these are modest values of $\rm G_{0}$,
\citet{anderson13} show that even with values of
external FUV radiation fields  as low as $\rm G_{0}=300$, 
external photoevaporation can
play a significant role
by greatly reducing the
lifetime of the disks, as well as truncating their outer edges.
This is consistent with the fact that 
a third of our disks have
radii $\leq$ 30 AU (see Table~\ref{tab:fit_par}).
As showed in \S\ref{sec:modelling} small
radii can reliably be determined from the SED.
These radii refer to the extent of the dust disk
since we lack resolved gas observations,
nevertheless, this result indicates that 
a possible explanation for the
presence of small disks in populated regions 
is due to external photoevaporation by massive OB stars. 

\begin{deluxetable}{lcc}
\centering
\tablewidth{0pt}
\tablecaption{KS test between histograms of spectral indices for the Taurus and 
\SOri samples \label{tab:KS_test}}
\tablehead{
\colhead{index}  & \colhead{D} & \colhead{p}
}
\startdata
$\rm n_{24-70}$ & 0.27 & 0.35 \\
$\rm n_{24-160}$ & 0.19 & 0.77 \\
$\rm n_{\rm K-24}$ & 0.14 & 0.96 \\
$\rm n_{\rm K-70}$ & 0.24 & 0.37 \\
$\rm n_{\rm K-160}$ & 0.29 & 0.19 \\
\enddata

\end{deluxetable}

Even though there is no correlation 
between disk radii and projected distance, a result
also found in the ONC proplyds \citep{vicente05,mann10,mann14},
our argument is reinforced by the fact that (a) 
two of our smallest disks, the sources SO697 ($R_{d}=7$ AU) 
and SO682 ($R_{d}=10$ AU), 
have the closest projected distances
to the \SOri multiple system 
(0.45 and 0.14 pc, respectively) 
(b) 81\% of our largest disks (e.g. the sources SO774, SO865 and SO1260) 
are located at
projected distances greater than 1 pc
and (c) Most of the 850 $\mu$m SCUBA-2
detections of \citet{williams13} are 
{\it outside} the PACS field, at distances
$\ge 2.23$ pc from the central ionizing sources. 

\begin{figure}[ht]
\includegraphics[width=1\columnwidth]{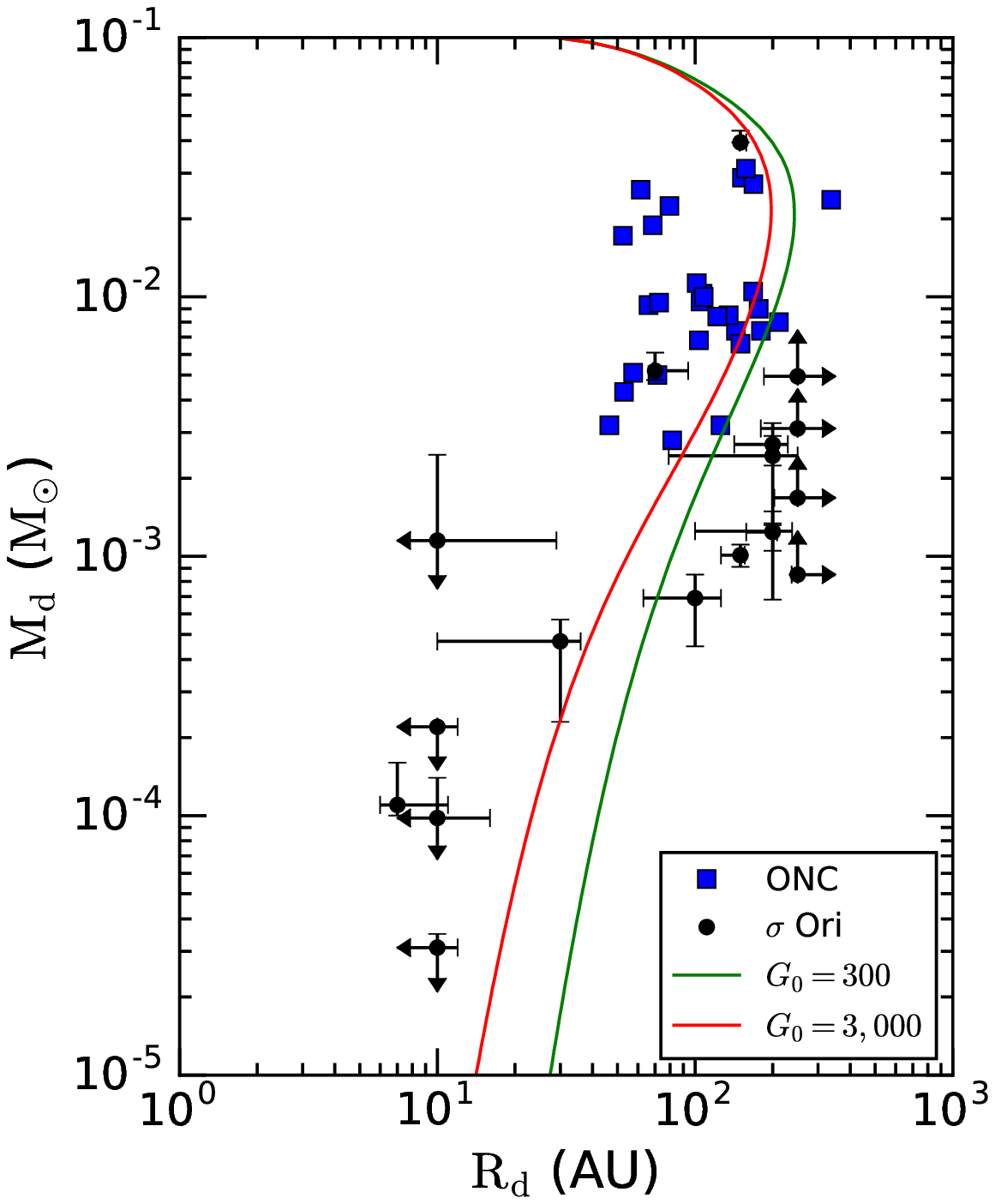}
\caption{Disk mass ($\rm M_{d}$) vs disk radius ($\rm R_{d}$) for \SOri sources (black dots) 
and proplyds in the ONC (blue squares). 
We have included the confidence intervals of $\rm M_{d}$ and $\rm R_{d}$
estimated in section~\ref{sec:modelling}.
The disk masses for the ONC are taken from \citet{mann10} and radii from \citet{vicente05}.
Evolutionary tracks (solid curves) are taken from \citet{anderson13} for two values of
$G_{0}$, $3000$ (red) and $300$ (green), and a viscosity parameter $\alpha= 0.01$.
}
\label{fig:Md_Rd_ONC}	 
\end{figure}

In Figure~\ref{fig:Md_Rd_ONC} we show our determinations of disk masses and radii
in \SOri compared to those of proplyds in the ONC.
We have included in the figure the confidence intervals of $\rm M_{d}$ and $\rm R_{d}$
estimated in section~\ref{sec:modelling}.
The disk masses for the ONC are taken from \citet{mann10} and
radii from \citet{vicente05}.
We also show the expected evolution of viscous disks
subject to photoevaporation from \citet{anderson13}.
Unlike isolated viscous evolution, in which disks expand as the
disk mass increases, viscous disks subject to external photoevaporation
first expand until they reach 
the evaporation radius
where photoevaporation begins to act and to rip out the external
disk regions. As a result, both the disk mass and radius
decrease with age \citep{anderson13}.
The \SOri disks fit this trend. Their masses are lower than
those of the ONC disks, since they are older, while their
radii are comparable or smaller. 
The two models shown in Figure~\ref{fig:Md_Rd_ONC} correspond to values
of  $G_0$, 300 and 3000, with the lower value corresponding to
overall larger disk radii.
This suggests that
the shift between 
the ONC disks and the \SOri disks may be due to the latter
being subject to comparatively lower levels of FUV flux (which
explains the slightly larger radii)
for a longer period of time (which may explain the smaller radii).
However, given that disk radii of proplyds were measured from HST/WFPC2 H$\alpha$ images, 
and hence these values are related to the gaseous disks, and we estimated disk radii from modeling the SEDs, 
which are associated with the dusty disks, 
there are likely systematic differences between the two 
disk radius measurements done using very different methods.
We need more sensitive
mm observations of gas on these disks in order to make a
more robust comparison.

The evolutionary models shown in Figure~\ref{fig:Md_Rd_ONC} assume a constant value of $G_0$ throughout the
disk evolution.
This in only an approximation since stars are not static, 
they orbit around the cluster center and therefore
will experience different FUV flux levels during their
lifetime. For very eccentric orbits the stars will stay longer in outer regions,
where the FUV field is low, retaining their disk masses 
longer. 
Furthermore, these models assume a constant value of alpha; 
adopting a non-uniform radial profile for the viscosity alters 
the disk structure \citep{kalyaan15}, and therefore the mass-radius evolutionary tracks.
We adopted a viscosity parameter of $\alpha$ = 0.01. 
With this viscosity the disks are depleted down to $1e-5$ $\rm M_{\odot}$ very quickly ($<$ 1Myr).
Using a lower $\alpha$ will increase the lifetime of the disks but will not 
produce disks as large as observed. 
Note that the timescale for disk dispersal depends strongly on the assumed disk viscosity, 
and therefore a relatively small range of alpha can lead to a wide range in the disk mass and radius at a given time.    
The models also assume an initial disk radius of 30 AU. In reality,
disks will exhibit a wide range of radii depending on the initial 
conditions of the core where the stars are formed. However, \citet{anderson13}
showed that varying the initial radius by a factor of two does not change
the evolutionary tracks. 
Another approximation of the models is assuming a stellar mass of $\rm M_{*}$ = 1 $\rm M_{\odot}$.
Stars with smaller masses will produce shallower gravitational potential wells, 
allowing the unbound material to be located closer to the star, and possibly producing smaller disks.
All these approximations may be responsible for the differences in radii observed between the models and 
disks in $\sigma$ Ori. 
Finally, as mentioned above, 
the radial distribution
of dust may differ from that of the gas.
However, even with these simplifying assumptions,
the models reproduce remarkably well the observations of the
ONC and qualitatively well the \SOri disks sample.

Additional support for the photoevaporative hypothesis
comes from the presence of forbidden lines in the optical spectra
of the disk sources. These lines are expected
to form in a photoevaporating wind \citep{hartigan95,acke05,pascucci09,
gorti11,rigliaco09,rigliaco12,rigliaco13,natta14}.
\citet{rigliaco09} found 
forbidden lines of [SII] and [NII] in high resolution spectra
of SO587, located at $d_p = 0.35$ pc. 
They associated these emission lines
with a photoevapotaring wind estimated with a $\rm \dot M_{loss} \sim$ 10$^{-9}$
$\rm M_{\odot}/yr$. Unfortunately, we did not model this object because
it was not detected with PACS.

\begin{figure}[ht]
\includegraphics[width=1\columnwidth]{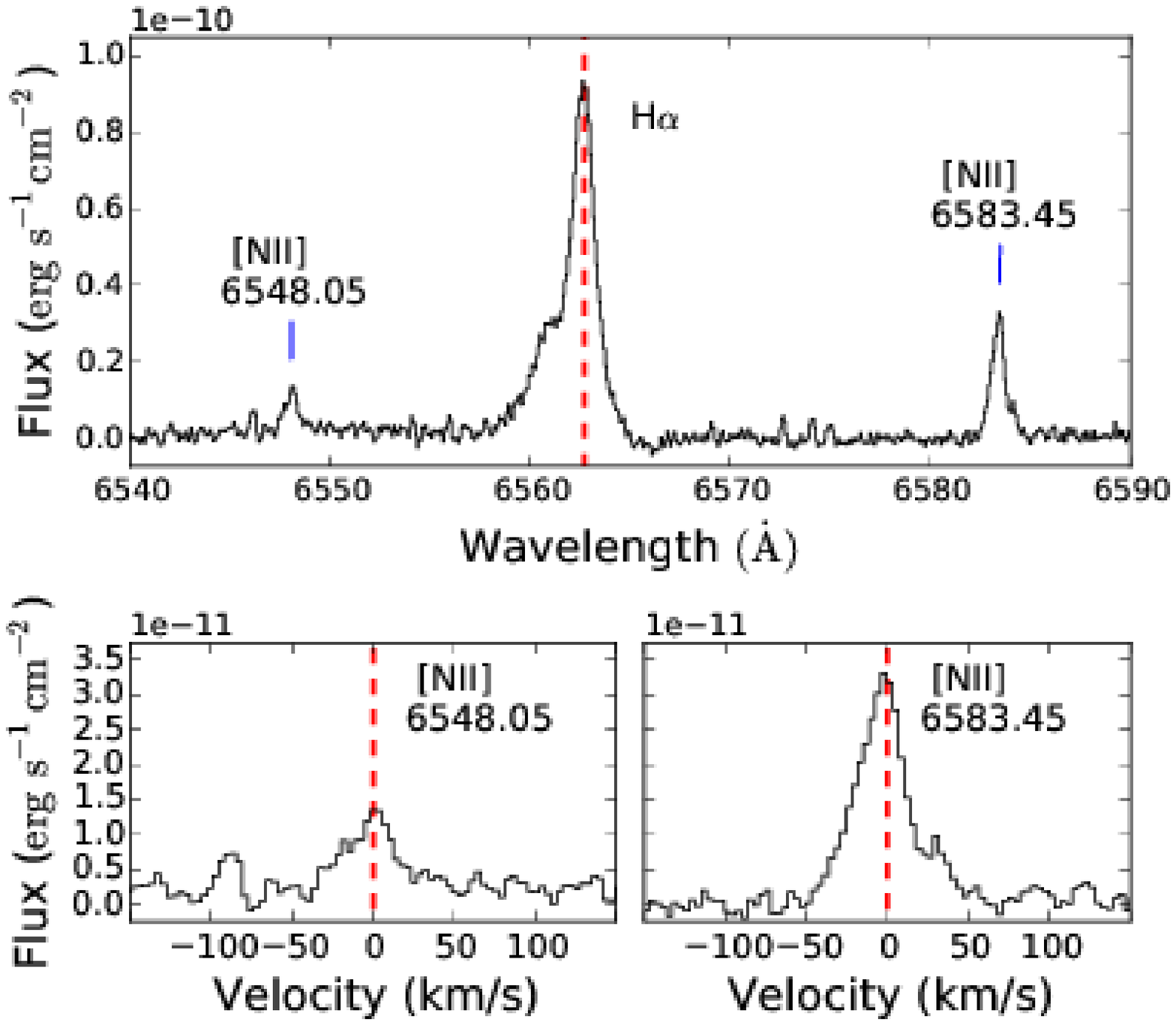}
\caption{{\it Top}: Hectochelle spectrum of SO662 (H14) showing the forbidden emission lines
of [NII] along with the $\rm H\alpha$ line characteristic of TTSs. 
{\it Bottom}: Lines of [NII] in velocity space.
This object is located at $\rm d_p$ = 0.60 pc 
and has a disk size $\rm R_d$ = 200 AU (Table~\ref{tab:fit_par}). 
The presence of the forbidden
lines of [NII] may be evidence of photoevaporation \citep{rigliaco09}.}
\label{fig:spectra1} 
\end{figure}

\begin{figure}[ht]
\includegraphics[width=1\columnwidth]{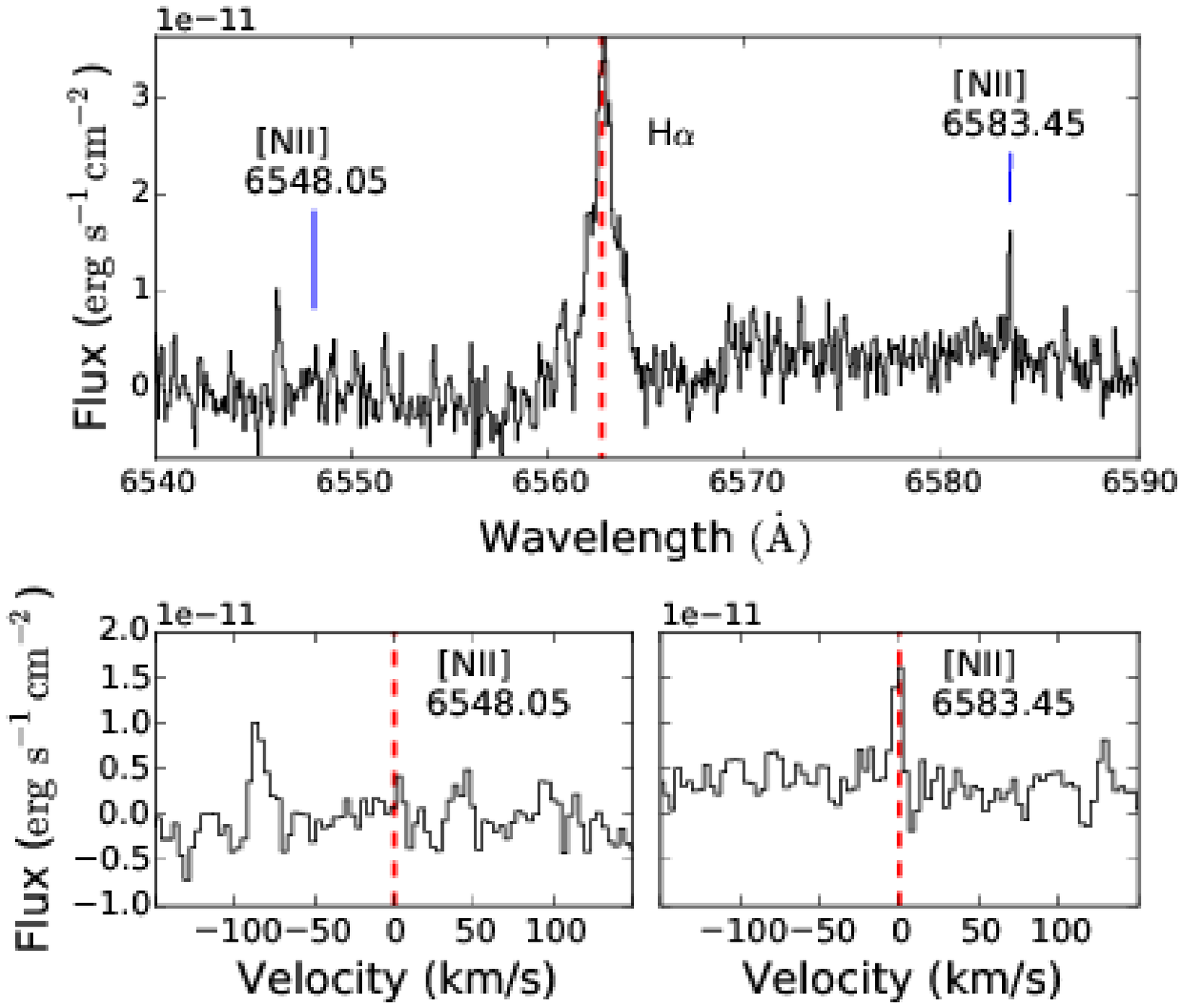}
\caption{{\it Top}: Hectochelle spectrum of SO697 (H14) showing the $\rm H\alpha$ line but
no emission lines of [NII]. {\it Bottom}: Spectrum in velocity space  
at the wavelenghts of the [NII] emission lines.
This object is the smallest disk in our sample with a disk size of $\rm R_d$ = 7 AU and located 
at $\rm d_p$ = 0.45 pc (Table~\ref{tab:fit_par}). 
The absence of the forbidden lines of [NII] may be the result of
the shrinking of the disk due to photoevaporation.}
\label{fig:spectra2} 
\end{figure}

The survey of H14 includes high resolution spectra
of some of the PACS sources;
these spectra cover
the region around H$\alpha$ and include the [NII] 
6548.05 and 6583.45 forbidden
lines.
Figure~\ref{fig:spectra1} displays on the top panel the spectrum 
of SO662 ($\rm R_d$ = 200 AU, $\rm M_d$ $\sim 2\, \rm M_{Jup}$; Table~\ref{tab:fit_par}) 
showing the H$\alpha$ line in emission, characteristic of
TTSs, and the forbidden lines of [NII].
This object is located at $d_{p}$ = 0.6 pc.
On the bottom panel
we also show the lines in velocity space. 
The blue-shift of these lines, if any, is smaller than
3 $\rm kms^{-1}$ (our velocity determination error)
consistent with the lowest blueshifts found in TTSs
for the [OI] line \citep{rigliaco13,natta14}. 
The low velocity value indicates 
that the forbidden line emission does not come from a jet
for which velocities range the hundreds of $\rm kms^{-1}$ \citep{hartigan95}.
Figure~\ref{fig:spectra2}, on the other hand, shows the spectrum of SO697, 
our smallest disk ($\rm R_d$ = 7 AU, $\rm M_d$ $\sim 0.1\, \rm M_{Jup}$; Table~\ref{tab:fit_par}) located at
$d_{p}$ = 0.45 pc. This object does not exhibit 
the [NII] forbidden line at 6548.05 $\rm \AA$ and 
has a very weak emission at 6583.45 $\rm \AA$ (bottom right panel). 
Similarly, none of our small disks 
with high resolution spectra
(SO396, SO462 and SO859) show any emission 
lines of [NII]. 

Since strong optical forbidden emission lines 
are expected to form in highly ionized regions,
as in the bright cusps characteristic of proplyds 
in the ONC,
the fact that
within the first 0.6 pc from $\sigma$ Ori,
one of our largest disks exhibits these lines while 
none of our small disks have any features,   
suggests that photoevaporation is in action in the cluster.
If this is indeed the case, then
SO662 is a photoevaporating disk in its 
initial phase while the small
disks are those that moved close enough to the hot stars during the lifetime of the cluster to be
stripped out of their outermost regions and hence cannot produce the forbidden lines.
We need a greater sample of disks with high resolution spectra 
covering a range of projected distances from the central stars
to confirm this hypothesis and characterize the region in the cluster
being affected by photoevaporation.

\section{Conclusions \label{sec:conclusions}}
\indent
We analyzed the IR emission of 32
TTSs (mostly class II stars) with PACS detections belonging to the \SOri cluster 
located in the Ori OB1b subassociation. 
We modeled 18 sources using the irradiated accretion disk
models of \citet{dalessio06}.
Our main conclusions are as follows:

\begin{enumerate}

	\item PACS detections are consistent with stars surrounded by optically
	thick disks with high 24-{\micron} excesses and spectral types between K2.0 and M5.0.

	\item Detailed modeling 
	indicates that
	most of our objects (60\%) can be explained by 
	$\epsilon$ = 0.01, indicative of significant 
	dust settling and possible grain growth. 
	This is consistent with
	previous studies of other young star-forming regions \citep{furlan09,mcclure10,manoj11}.

	\item
	61\% of our disks can be modeled with large sizes (R$_{\rm d} \geq$ 100 AU).
	The rest,
	have dust disk radii of less than 80 AU. 
	These disks may have been subject
	to photoevaporation. We estimated that 80\% of our disks are exposed to
	FUV fluxes between 300 $\lesssim$ G$_{0} \lesssim$ 1000. These values
	may be high enough to photoevaporate the outer edges of the closer disks.
	Additionally, within the first 0.6 pc from the central ionizing sources 
	we found forbidden emission lines of [NII] in SO662 ($\rm R_d = 200$ AU) while none of the small disks
	exhibit any features. This suggests that 
	the region producing the lines is located in the outer disk. Therefore, 
	SO662 may be a photoevaporative disk in its initial phase while the small disks have already 
	photoavaporated most of their material and hence cannot produce the [NII] lines. 

	\item The masses of our disks range between 0.03 to $\sim$39 M$_{\rm Jup}$,
	with 35\% of the disks having masses 
	lower than 0.001 M$_{\odot}$, i.e., 1 Jupiter mass. These low
	masses suggest that the formation of giant planets 
	is probably over in the cluster. If this is the case, then 
	time scales for giant planet formation should be less than
	3 Myr, or giant planets are difficult to form in clustered environments.

\end{enumerate}

This work was supported by UNAM-PAPIIT grant number IN110816 to JBP. 
Model calculations were performed in the supercomputer Miztli, at 
DGTIC-UNAM. KM acknowledges a scholarship from CONACYT 
and financial support from CONACyT grant number 168251. JH acknowledges 
UNAM-DGAPA's PREI program for visiting research at IRyA.  We have made 
extensive use of NASA-ADS database.

\end{document}